\documentstyle[12pt,axodraw]{article}

\input{psfig}

\def\pict #1 by #2 (#3){
  \vbox to #2{
    \hrule width #1 height 0pt depth 0pt
    \vfill
    \special{picture #3} 
    }
  }

\def\scaledpicture #1 by #2 (#3 scaled #4){{
  \dimen0=#1 \dimen1=#2
  \divide\dimen0 by 1000 \multiply\dimen0 by #4
  \divide\dimen1 by 1000 \multiply\dimen1 by #4
  \picture \dimen0 by \dimen1 (#3 scaled #4)}
  }

\def\F2{\pict 3.00in by 2.00in (F2)}
\def\Q2nuplane{\pict 3.00in by 2.00in (Q2nuplane)}

\def\pq{p \! \cdot \! q}
\def\bar{\overline}
\def\Tr{\mbox{Tr}}

\def\beq{\begin{equation}}
\def\eeq{\end{equation}}
\def\ni{\noindent}

\newcommand{\ls}[1]
   {\dimen0=\fontdimen6\the\font 
    \lineskip=#1\dimen0
    \advance\lineskip.5\fontdimen5\the\font
    \advance\lineskip-\dimen0
    \lineskiplimit=.9\lineskip
    \baselineskip=\lineskip
    \advance\baselineskip\dimen0
    \normallineskip\lineskip
    \normallineskiplimit\lineskiplimit
    \normalbaselineskip\baselineskip
    \ignorespaces
   }

   {\arrayclosep 0.14em\begin{eqnarray}}{\end{eqnarray}}

\begin{document}

\title{The Electromagnetic Mass Differences of Pions and 
Kaons}

\author{John F. Donoghue$^{(a)}$ and Antonio F. 
P\'{e}rez$^{(a, b)}$\\[5mm]
(a) Department of Physics and Astronomy \\
University of Massachusetts, Amherst, MA~01003 \\
(b) Department of Physics \\
University of Cincinnati, Cincinnati, OH~45221}

\date{}

\begin{titlepage}
\maketitle
\begin{abstract}
\thispagestyle{empty}

We use the Cottingham method to calculate the pion and kaon
electromagnetic mass differences with as few model dependent inputs as
possible.  The constraints of chiral symmetry at low energy, QCD at
high energy and experimental data in between are used in the
dispersion relation.  We find excellent agreement with experiment for
the pion mass difference.  The kaon mass difference exhibits a strong
violation of the lowest order prediction of Dashen's theorem, in
qualitative agreement with several other recent calculations.

\end{abstract}
{\vfill 
UCTP-9-96\\
UMHEP-428\\
hep-ph/9611331}
\end{titlepage}

\noindent
{\large{\bf 1. Introduction}}
\bigskip
\bigskip

The calculation of the electromagnetic mass differences of pions and
kaons has recently been quite an active topic
\cite{Bard:89}-\cite{Dono:93a} in the field of chiral
perturbation studies.  This is partially due to the interest in the
values of the light quark mass ratios, for which we need to be able to
separate electromagnetic from quark mass effects
\cite{Dono:94b}-\cite{Leut:96}. The realization that Dashen's
theorem \cite{Dash:69}, relating pion and kaon electromagnetic mass
differences in the limit of vanishing $m_u, m_d, m_s $, could be
significantly violated in the real world
\cite{Bijn:93,Bijn:96,Dono:93a,Dunc:96} has added to the importance of
a direct calculation of these electromagnetic effects.  Moreover these
calculations have an intrinsic interest as state of the art
investigations of our ability to handle new types of chiral
calculations.  The classic studies of chiral perturbation theory
\cite{Gass:84,Gass:85a} are being extended to calculations where one
must obtain more detailed information of the intermediate energy
region using dispersion relations (or sometimes models).  The
electromagnetic mass differences are nonleptonic amplitudes which are
a challenge to calculate in a controlled fashion.  It is our goal in
this paper to calculate these mass differences as well as we can at
present.

Our tool is the Cottingham method for calculating electromagnetic mass
differences.  As explained more fully in Section 3, this converts the
mass differences amplitude into a dispersion integral over the
amplitudes for $\gamma \pi$ inelastic scattering.  As we learned in
the 1970's from the study of $\gamma p$ inelastic scattering, the
physics of such a process is reasonably simple.  The elastic
scattering is well known.  At low energies, one sees the inelastic
production of the low lying resonances.  In our study we take these
resonances and their coupling constants from experimental data.  At
high energies one enters the deep inelastic region for which
perturbative QCD can be used.  It turns out that in the pion mass
difference the deep inelastic region cancels out both at zeroth and
first order in the quark masses.  This leaves the mass differences to
be dominated by the lower energy region.

There are a series of constraints on the calculation which are 
important for giving us control 
over our method and results.  The most important of these 
are:

\begin{enumerate}
\item There exists a rigorous result for these mass 
differences, exact in the limit that $m_q 
\rightarrow 0, (q = u,d)$ which states that in this chiral limit 
the pion mass difference is

\begin{equation}
\Delta m^2_{\pi} = -{3\alpha \over 4\pi F_\pi^2}\int ds s ln s 
(\rho_V(s) - \rho_A(s)) \ \ ,
\end{equation}

\noindent where $\rho_V(s)$ and $\rho_A(s)$ are the vector 
and axial vector spectral functions measured in $e^+ e^-$ annihilation
and in $\tau$ decays \cite{Dono:94a}.  This is a powerful constraint
because it requires that the full calculation differ from this only by
terms of order $m^2_{\pi}$ or higher, and must reduce to this as
$m^2_{\pi}
\rightarrow 0$.  Many of these deviations are kinematic in origin and
hence are well tied down by this constraint.

\item Dashen's theorem states that in an $SU(3)$ extensions 
of this same limit $m_q = 0, q 
= u, d, s)$ that the kaon mass difference is equal to the pion 
mass difference.  This means 
that similar physics enters both amplitudes and one is able to 
focus more directly on 
$SU(3)$ breaking.

\item The low energy structure of the Compton amplitudes 
$\gamma \pi \rightarrow \gamma 
\pi$ and $\gamma K \rightarrow \gamma K$ are known 
rigorously from chiral perturbation theory \cite{Bijn:88}-\cite{Dono:93b} and
the process in the crossed channel $\gamma
\gamma
\rightarrow \pi \pi$ matches well with experiment \cite{Dono:93b}.

\item QCD gives us important information about the high 
energy behavior of the dispersive 
integral, with the result that $\Delta m^2_{\pi}$ is finite up 
to order $m^2_q$, while 
$\Delta m^2_K$ has at most a logarithmic divergence at 
order $m_q$, which is to be 
absorbed into the $u, d$ quark masses.  This is very useful 
in pinning down the high 
energy parts of the calculation.

\item The medium energy intermediate states are known 
directly from experiment.  This 
region is the most difficult to control purely theoretically, 
and so we rely on experimental 
data to overcome our inability to provide a first-principles 
theoretical calculation.

\end{enumerate}

\noindent These properties are important ingredients for the 
reliability of our method.  
While there are still some approximations and educated 
guesses involved in the matching 
up of the various regions of the calculation, this method is 
more than just another model 
and represents the real world as well as is possible in 
analytic calculations at present.

While we estimate that our uncertainty is about 10\% for pions, and
20\% for kaons, our calculated value for the pion mass difference
agrees excellently with experiment $(\Delta m^{th}_{\pi} = 4.54 \pm
0.50$ MeV vs. $\Delta m^{expt}_{\pi} = 4.60$ MeV).  In the case of
kaons, our calculated value is $\Delta m^{th}_{K} = 2.6 \pm 0.6$ MeV,
indicating a strong breaking of Dashen's theorem $(\Delta m^{DT}_K =
1.3 MeV)$ in agreement with many other recent works
\cite{Bijn:93}-\cite{Dono:93a}.

In the next section, we briefly review the physics and history 
of the calculations of 
electromagnetic mass differences.  Section 3 presents the 
basics of the Cottingham method, 
while Section 4 describes our application of it to the pion 
mass difference.  The kaon mass 
difference is studied in Section 5, and we summarize our 
findings in Section 6.

\bigskip
\bigskip
\noindent
{\large{\bf 2. Review of the Problem}}
\bigskip
\bigskip

The mass differences of kaons and pions

\begin{eqnarray}
\Delta m^{expt}_{\pi} & \equiv & m_{\pi^{\pm}} - 
m_{\pi^0} \ = \ 4.5936 \pm 0.0005 \ \ \mbox{MeV} \ \ ,
\nonumber \\
\Delta m^{expt}_K & \equiv & m_{K^{\pm}} - m_{K^0} = 
-3.995 \pm 0.0034 \ \ \mbox{MeV} \ \ ,
\end{eqnarray}

\noindent or

\begin{eqnarray}
\Delta m^2_{\pi} & = & 2 m_\pi^{(avg)} \Delta m_\pi \ \ = \ \ 
(1.2612 \pm 0.0001) \ 10^{-3} \ \ \mbox{GeV}^2 \ \ ,
\nonumber \\
\Delta m^2_{K} & = & 2 m_K^{(avg)} \Delta m_K \ \ = \ \  
(-3.9604 \pm 0.0035) \ 10^{-3} \ \ \mbox{GeV}^2 \ \ , 
\end{eqnarray}

\noindent 
where 
$ m_{K,\pi}^{(avg)} \equiv {{1}\over{2}} (m_{(K,\pi)^\pm} +  m_{(K,\pi)^0} )$,
are due to two sources:  quark masses and 
electromagnetic interactions.  The 
difference in mass of the up and down quarks can produce 
isospin breaking in hadron 
masses.  However, because the quark mass splitting is 
$\Delta I = 1$ and the pion mass 
difference is only sensitive to $\Delta I = 2$ effects, the pion 
mass difference only receives 
contributions of second order, i.e., $(m_d - m_u)^2$.  In 
fact the leading effect of this 
order is calculable in chiral perturbation theory

\begin{equation}
\left. \Delta m_{\pi}^2 \right)_{QM}  \ \ = \ \ 
{1 \over 4}  {{(m_u - m_d)^2} \over {(m_u+m_d)(m_s - \hat{m})}} 
m^2_{\pi^\pm}
\ \ ,
\end{equation}

\noindent and is quite small.  To the level of our 
approximations we will neglect this quark 
mass effect and treat the pion mass difference as purely 
electromagnetic.  The kaon mass 
difference, on the other hand, does receive an important 
contribution linear in $m_d - m_u$

\begin{equation}
\left.  
\Delta m^2_K \right)_{QM} = {m_u - m_d \over m_u 
+ m_d} m^2_{\pi} + {\em O} 
\left( (m_u - m_d)^2 \right) \ \ .
\end{equation}

\noindent This relation is one of the primary sources of 
information on quark mass ratios.  
For it to be useful we need to known how much of the kaon 
mass difference is due to 
electromagnetic interactions.

We have one handle on the electromagnetic mass differences which comes
purely from symmetry considerations.  The electromagnetic interaction
explicitly violates chiral SU(3) symmetry, and its effect can be
described within the chiral energy expansion.  At lowest order, which
is order $p^0$, the unique effective Lagrangian with the right
symmetry breaking properties is

\begin{equation}
{\cal L}_0 = g_{EM} \: Tr(QUQU^+) \ \ .
\end{equation}

\noindent This Lagrangian produces no shift in the masses of 
neutral mesons and equal shifts for $\pi^+$ and $K^+$, so that it
results in

\begin{equation} 
\Delta m^2_{\pi} = \Delta m^2_K \ \ .
\end{equation}

\noindent 
This equality is known in the literature as Dashen's theorem
\cite{Dash:69}. It is valid in the limit of vanishing quark masses
(u, d and s) and hence of massless pions and kaons.  There are a large
number of effective Lagrangians possible with extra derivatives and/or
factors of the quark masses, so that Dashen's theorem will receive
corrections of order $m_s$ or equivalently of order $m^2_K$
\cite{Urec:95}.  Unfortunately the coefficients of the higher order
Lagrangians are not known, so that one cannot obtain the corrections
to Dashen's theorem from symmetry considerations.  A direct
calculation is required.

In order to obtain the electromagnetic mass shifts, one must 
calculate

\begin{equation}
\delta m^2 = {ie^2 \over 2} \int d^4 x \langle \pi (p) \left| T 
J_{\mu} (x) J_\nu (0) \right| \pi 
(p) \rangle D^{\mu \nu}_F (x)
\end{equation}

\noindent as in Figure 1.

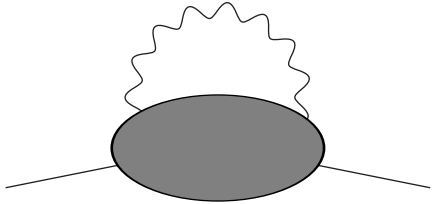
\begin{figure}[h]
\ls{1.0}
\begin{center}
\begin{picture}(210,120)(0,0)
\Line(30,40)(80,50)
\Line(140,50)(190,40)
\PhotonArc(110,75)(32,340,200) 3 9
\GOval(110,55)(20,40)(0) {.5}
\end{picture}
\end{center}
\caption[Em self-energy.]
{\label{fig:sefd} \sf Electromagnetic self-energy.}
\end{figure}

\ni
In momentum space this is

\begin{equation}
\delta m^2 = {ie^2 \over 2} \int {d^4q \over (2 \pi)^4} 
{g^{\mu \nu} T_{\mu \nu} (q^2, p 
\cdot q) \over q^2 + i \epsilon} \ \ ,
\label{dmdirect}
\end{equation}

\noindent where 

\begin{equation}
T_{\mu \nu} (q^2, p \cdot q) = i \int d^4x e^{-iq \cdot x} 
\langle \pi (p) \left| T J_{\mu} 
(x) J_{\nu} (0) \right| \pi (p) \rangle \ \ .
\end{equation}

\noindent This calculation is different from standard 
calculations within chiral perturbation 
theory, because we need to be able to explicitly calculate 
(and not just parametrize) the 
medium energy and high energy contributions.

There are a few things that we know rigorously about the calculation.
Within QCD, the high energy renormalization of a quark mass involves a
logarithmic divergence which is proportional to the quark mass itself.
Therefore the pion mass difference can pick up divergence's only
proportional to the second power of the quark masses, which will go
into defining renormalized masses in Eq. (4).  In our approximation,
or more strictly in the chiral limit, the pion electromagnetic mass
difference is finite.  For the kaon there may appear a divergence of
order $\alpha m_u$ or $\alpha m_d$, i.e., suppressed by one power of
the light quark masses.  This goes into a renormalization of the quark
masses in Eq. (5).  In principle there is an ambiguity about how much
of the electromagnetic interaction goes into the renormalized values
of the quark masses.  This can only be solved by a precise
renormalization condition which defines the renormalized quark masses.
However because this ambiguity is proportional to $\alpha m_u$ and
$\alpha m_d$, while the kaon mass difference needs only one factor of
$\alpha$ {\em or} $(m_d - m_u)$, this ambiguity is tiny and is far
below the sensitivity of our calculation.

The earliest attempts at explicit calculations (Riazuddin
\cite{Riaz:59} and Socolow \cite{Soco:65})
appeared plausible but can now be recognized as mistreating the chiral
portions of the calculation.  The earliest valid method, and still a
remarkably beautiful result, came in the work of Das et. al.
\cite{Das:67}.  Here soft pion theorems were used to turn the matrix
element of Eq.~(10) into a vacuum polarization function, which in turn
can be written as a dispersion relation in terms of the spectral
functions of the vector and axial vector currents, yielding the
formula quoted in Eq.~(1).  Since QCD satisfies the chiral and high
energy properties assumed in the original derivations, this remains an
exact statement of QCD in the limit $m_u = m_d = 0$.  The original
authors saturated the spectral functions by a single vector and axial
vector pole satisfying the Weinberg sum rules \cite{Wein:67}, leading
to a remarkably good value $\Delta m^{Das}_{\pi} = 5.0$ MeV.  More
recently this sum rule has been explored using the measured spectral
functions from $e^+ e^-$ annihilation and $\tau$ decay, plus QCD
constraints
\cite{Dono:94a}. These show that the physics of the pion mass
difference is remarkably simple in the chiral limit with the most
important effects being those of the lightest resonance contributions.
The Das et. al. calculation remains a benchmark for other calculations
and will be an important constraint on our work.

Through the experience of the past decade of studies of chiral
perturbation theory, we have gained some insight into the physics of
intermediate energies.  This lead to model attempts to calculate
electromagnetic mass differences \cite{Dono:93a}. These calculations
showed a large breaking (up to a factor of 2) of Dashen's theorem due
to mass effects.  To a large extent the violation of Dashen's theorem
has a simple kinematic origin in the pseudoscalar propagators of the
one loop diagram.\footnote{Ref. \cite{Dono:93a} has an error in one of
the mass effects, as described later.  We disagree with the
methodology of a paper which attempted to correct this problem
\cite{Baur:96}, and agree with the critique of \cite{Baur:96} which is
contained in \cite{Bijn:96}.} Lattice simulations have also recently
started to be applied to this problem.  They also see a significant
violation of Dashen's theorem, ( $\Delta m_K = 1.9$ MeV )
\cite{Dunc:96}.

\bigskip
\bigskip
\noindent
{\large{\bf 3. The Cottingham method and meson mass shift}}
\bigskip
\bigskip

The nonleptonic matrix element which we must calculate is given in Eq.
(10).  If we decompose the Compton amplitude in terms of gauge
invariant tensors, we can define

\begin{eqnarray}
T_{\mu \nu} (q^2, p \cdot q) & = & D_{1 \mu \nu} T_1 
(q^2, p \cdot q) + D_{2 \mu \nu} T_2 (q^2,
p \cdot q) \ \ ,
\nonumber \\
D_{1 \mu \nu} & = & - g_{\mu \nu} + {q_{\mu} q_{\nu} 
\over q^2} \ \ ,
\nonumber \\
D_{2 \mu \nu} & = & {{1}\over{p^2}}
\left( p_{\mu} - {p \cdot q \over q^2} 
q_{\mu} \right) \, \left( p_{\nu} -
{p \cdot q \over q^2} q_{\nu} \right) \ \ .
\end{eqnarray}

\ni
We have used the standard definitions for these tensors. Note
that in the soft-pion limit, i.e. $p_\mu \rightarrow 0$, the combination
$D_{2\mu\nu} T_2$ vanishes as we will see in the following section.


A first step consists of a rotation in the complex plane and a change
of variables. We work in the pion rest frame, $p \cdot q = m_{\pi}
q_0$.  Since the singularities in $T_{\mu \nu}$ are located just below
the positive real axis and above the negative real axis in the complex
$q_0$ plane, the integration over $q_0$ may be rotated to the
imaginary axis, $q_0 \rightarrow iq_0$, without encountering any
singularities.  After this transformation the integral involves only
space\-like moments for photons, i.e., $q^2 \equiv - Q^2 = - \left(
q^2_0 + \vec{q}^{\; 2} \right)$, and the mass shift becomes

\begin{equation}
\delta m^2 = 
{e^2 \over 2} \int {d^3 \vec{q} \: dq_0 \over (2 \pi)^4} \: 
{{g^{\mu \nu} T_{\mu \nu} (\vec{q}, iq_0)} 
\over {q^2_0 + {\vec{q}}^{\; 2}}} \ \ .
\end{equation}

\noindent 
A change of variables from $(\vec{q}, q_0)$ to 
$(Q^2, \nu)$, where $\nu = m_\pi q_0$, involves

\begin{equation}
\int d^3 \vec{q} \: dq_0 = 
2 \pi \int_{0}^{\infty} dQ^2 
\int_{m_\pi Q}^{- m_\pi Q} 
{{d \nu}\over{ m_\pi^2}} 
\sqrt{ m_\pi^2 Q^2 - \nu^2} \ \ ,
\end{equation}

\noindent which converts the mass shift to

\begin{eqnarray}
\delta m^2 & = & 
{e^2 \over 16 \pi^3} \int^{\infty}_{0} 
dQ^2 \int^{m_\pi Q}_{- m_\pi Q} {{d \nu} \over {m_\pi^2}} 
{\sqrt{ m_\pi^2 Q^2 - \nu^2} \over Q^2} 
g^{\mu \nu} T_{\mu \nu} (-Q^2, i \nu) 
\nonumber \\
& = & 
{e^2 \over 16 \pi^3} \int^{\infty}_{0} 
dQ^2 \int^{m_\pi Q}_{- m_\pi Q} {{d \nu} \over {m_\pi^2}} 
{\sqrt{ m_\pi^2 Q^2 - \nu^2} \over Q^2} 
\nonumber \\
& + & \left[ -3 T_1 (-Q^2, i \nu) + 
  \left( 1 - {{\nu^2} \over {m_\pi^2 Q^2}} \right) 
  T_2 (-Q^2, i \nu) \right] \ \ .
\end{eqnarray}

\noindent This has reduced the mass shift to an integral over 
the forward Compton scattering
amplitude for space-like photons.

The reduced Compton amplitudes $T_1$ and $T_2$ are 
presently required to be evaluated at
imaginary momenta, $i \nu$.  However they can be written in 
terms of physical amplitudes via
dispersion relations.  The Compton amplitudes are known to 
obey dispersion relations in the
$\nu$ variable with that for $T_1$ requiring one 
subtraction.

\begin{eqnarray}
T_1 (q^2, \nu) &=& T_1 (q^2, 0) + {\nu^2 \over \pi} \, 
\int^{\infty}_{0} \! \! {d {\nu^\prime}^2 \over
{\nu^\prime}^{2}} \: 
{Im T_1 (q^2, \nu^\prime) \over { {\nu^\prime}^2 - \nu^2}} \ \ ,
\nonumber \\
T_2 (q^2, \nu) &=& {1 \over \pi} \, \int^{\infty}_{0} \! \! \! \! 
d {\nu^\prime}^2 \, { {Im T_2 (q^2, \nu^\prime)} 
\over { {\nu^\prime}^2 - \nu^2}} \ \ .
\end{eqnarray}

The imaginary part of the forward scattering amplitudes $Im 
T_i$ are defined as electron scattering structure functions

\begin{equation}
{1 \over \pi} Im T_i ( -Q^2, \nu) = W_i (-Q^2, \nu ) 
\ \ , \ \ \mbox{for} \ i = 1,2 \ \ .
\end{equation}

\noindent 
After employing these dispersion relations, the integral over $\nu$
can be done explicitly with the result

\begin{eqnarray}
\Delta m^2 &=& \! {\alpha \over 4 \pi} \int^{\infty}_{0} \!   
 {dQ^2 \over Q^2} \, \left\{ 
  - {3 \over2} Q^2 T_1 (-Q^2, 0) 
 \phantom{\Lambda_1 \left( {{{\nu^\prime}^2 } \over { m_\pi^2 Q^2}} \right)}
 \right. 
\nonumber \\
&+& \! 3Q^2 \int^{\infty}_{0} \! 
 { { {d \nu^\prime}^2 } \over {{\nu^\prime}^2 } } \, 
 W_1 (-Q^2, \nu^\prime) \, 
 \Lambda_1 \left( { { {\nu^\prime}^2 } \over { m_\pi^2 Q^2}} \right)
\nonumber \\
&+& \! \left.  \int^{\infty}_{0} 
{ { {d \nu^\prime}^2 } \over {m_\pi^2} } \,
 W_2 (-Q^2, \nu^\prime) 
\Lambda_2 \left( {{\nu^\prime}^2 \over { m_\pi^2 Q^2}} \right)
\right\} \ \ ,
\label{dmcott}
\end{eqnarray}

where

{\everymath{\displaystyle}
\begin{eqnarray}
\Lambda_1(y) & \equiv & {{1}\over{2}} +y -y \sqrt{ 1 + {{1}\over{y}} } \ \ ,
\nonumber \\
\Lambda_2(y) & \equiv & 
-{{3}\over{2}} -y +(1 + y) \sqrt{ 1 + {{1}\over{y}}} \ \ .
\label{defL1L2}
\end{eqnarray}
}

\noindent 
These manipulations have transformed the mass shifts into integrals
over the structure functions in the physical region, as well as the
subtraction term $T_1 (-Q^2, 0)$.  The $(Q^2, \nu)$ plane is shown in
Fig. 3.1, as is the physical region where $W_i \neq 0$.



\bigskip
\bigskip
\noindent
{\large{\bf 4. The pion EM mass difference}}
\bigskip
\bigskip

   In this section, we describe the details of the calculation of the
electromagnetic mass difference of the pion. The exact result of the
Das et al. \cite{Das:67} calculation in the chiral limit involves the
difference of spectral functions $\rho_V(s) - \rho_A(s)$. This
difference is entirely determined by the leading vector and
axial-vector resonances.  Therefore, our first step is to study the
low energy chiral amplitudes supplemented by the interactions of
vector and axial-vector resonances. These have been previously studied
in a model field theoretic calculation \cite{Dono:93a}. We correct a
technical mistake in that work (which was also noted in
\cite{Baur:96}), and transform the results into our dispersive
framework. This allows us to show how the Cottingham method merges
with the chiral limit result of Das et al. as $m_\pi \to 0$.

We subsequently generalize the calculation by treating the resonances
more realistically and adding in other ingredients to the amplitude.
The former improvement involves the replacement of the
``narrow-width'' treatment of the resonances, which occurs in any
field theoretic treatment, by spectral functions which account for the
energy variation and width of the resonances. To complete the
ingredients to the calculation, we add resonance transitions not
accounted for previously and also the deep inelastic continuum. The
resonance couplings follow from experiment, and their presence in the
Compton amplitude is confirmed by the comparison of theory and
experiment in $\gamma\gamma \to \pi^0\pi^0$
\cite{Bijn:88,Dono:88,Dono:93b,Ko:90}. The deep inelastic region cancels in
the mass difference to the order that we are working, so that we
include only a few comments on the matching of low and high energies.

\bigskip
\bigskip
\ni
{\it Lagrangian with spin-1 resonances}
\bigskip

Our starting point for calculating the pion Compton scattering
amplitude is a Lagrangian which includes ${\cal O} (E^2)$ chiral terms
and ${\cal O} (E^4)$ vector ($J^{PC} =1^{--}$) and axial-vector
($J^{PC} = 1^{++}$) couplings, introduced by Ecker et
al.~\cite{Ecke:89a,Ecke:89b}. This Lagrangian provides an accurate description
at low and medium energies (up to $\sim 1$ GeV).

{\everymath{\displaystyle}
\begin{eqnarray}
{\cal L} & = & - {{1}\over{4}} F_{\mu\nu} F^{\mu\nu} +
{{F_\pi^2}\over{4}} \Tr \left( D_\mu U D^\mu U^\dagger + \chi
U^\dagger + \chi^\dagger U \right)
\nonumber \\
& & - {{1}\over{2}} \Tr \left( \nabla^\lambda V_{\lambda\nu}
\nabla_\nu V^{\nu\mu} - {{1}\over{2}} M_V^2 V^{\mu\nu} V_{\mu\nu}
\right)
\nonumber \\
& & + {{F_V}\over{2 \sqrt{2}}} \Tr \left( V_{\mu\nu} f_+^{\mu\nu}
\right) + {{i G_V}\over{\sqrt{2}}} \Tr \left( V_{\mu\nu} u^\mu u^\nu
\right)
\nonumber \\
& & - {{1}\over{2}} \Tr \left( \nabla^\lambda A_{\lambda\nu}
\nabla_\nu A^{\nu\mu} - {{1}\over{2}} M_A^2 A^{\mu\nu} A_{\mu\nu}
\right)
\nonumber \\
& & + {{F_A}\over{2 \sqrt{2}}} \Tr \left( A_{\mu\nu} f_-^{\mu\nu}
\right)
\ \ ,
\label{Lchires}
\end{eqnarray}
}

\noindent
The notation is defined in the appendix. The relevant terms after expanding the
above Lagrangian in terms of pion, photon and spin-1 resonance fields are

{\everymath{\displaystyle}
\begin{eqnarray}
{\cal L} & = & 
i e A^\mu ( \pi^+ \partial_\mu \pi^-  - \pi^- \partial_\mu \pi^+ )
+ e^2 A^\mu A_\mu \pi^+ \pi^- 
\nonumber \\
& & - {{e F_V}\over{2}} F^{\mu\nu} \rho_{\mu\nu}^0 
 \left( 1 - {{\pi^+ \pi^-}\over{F_\pi^2}} \right)
\nonumber \\
& & + {{i G_V}\over{F_\pi^2}} \rho_{\mu\nu}^0 
 ( \partial^\mu \pi^+ \partial^\nu \pi^- 
 + \partial^\mu \pi^- \partial^\nu \pi^+ ) 
\nonumber \\
& & - {{2 e G_V}\over{F_\pi^2}} A^\mu \rho_{\mu\nu}^0 
 ( \pi^+ \partial^\nu \pi^- + \pi^- \partial^\nu \pi^+ ) 
\nonumber \\
& & - {{i e F_A}\over{2 F_\pi}} F^{\mu\nu}  
 \left( a_{1_{\mu\nu}}^- \pi^+ - a_{1_{\mu\nu}}^+ \pi^- \right) \ \ .
\label{Lvector}
\end{eqnarray}
}

\begin{figure}[htbp]
\ls{1.0}
\begin{center}


\begin{picture}(360,80)(0,0)

\Line(40,30)(80,30)
\Line(80,30)(120,30)
\Line(120,30)(160,30)
\Photon(40,50)(79,32)2 4
\Photon(121,31)(160,50)2 4
\GCirc(80,30)3 1
\GCirc(120,30)3 1

\Line(200,30)(240,30)
\Line(240,30)(280,30)
\Line(280,30)(320,30)
\Photon(241,31)(320,50)2 6
\Photon(200,50)(278,32)2 6
\GCirc(240,30)3 1
\GCirc(280,30)3 1

\put(170,10){(a)}

\end{picture}

\begin{picture}(360,70)(0,0)

\Line(120,30)(180,30)
\Line(180,30)(240,30)
\Vertex(180,30)2 
\Photon(180,30)(120,50)2 5
\Photon(180,30)(240,50)2 5

\put(170,10){(b)}

\end{picture}

\begin{picture}(360,70)(0,0)

\Line(40,30)(100,30)
\Line(100,30)(160,30)
\Photon(70,40)(40,50)2 2
\Line(100,29)(70,39)
\Line(100,31)(70,41)
\Vertex(70,40)2
\Photon(100,30)(160,50)2 5
\Vertex(100,30)2

\Line(200,30)(260,30)
\Line(260,30)(320,30)
\Photon(260,30)(200,50)2 5
\Photon(290,40)(320,50)2 2
\Line(260,29)(290,39)
\Line(260,31)(290,41)
\Vertex(290,40)2
\Vertex(260,30)2

\put(170,10){(c)}

\end{picture}

\begin{picture}(360,70)(0,0)

\Line(40,30)(80,30)
\Line(80,30)(120,30)
\Line(80,29)(120,29)
\Line(80,31)(120,31)
\Line(120,30)(160,30)
\Photon(40,50)(79,30)2 4
\Photon(121,30)(160,50)2 4
\Vertex(80,30)2
\Vertex(120,30)2

\Line(200,30)(240,30)
\Line(240,30)(280,30)
\Line(240,29)(280,29)
\Line(240,31)(280,31)
\Line(280,30)(320,30)
\Photon(241,30)(320,50)2 6
\Photon(200,50)(278,30)2 6
\Vertex(240,30)2
\Vertex(280,30)2

\put(170,10){(d)}

\end{picture}


\begin{picture}(360,80)(0,0)

\Line(20,30)(60,30)
\Line(60,30)(100,30)
\Photon(60,30)(60,70)2 4
\GCirc(60,30)3 1 

\Line(115,28)(125,28)
\Line(115,32)(125,32)

\Line(140,30)(180,30)
\Line(180,30)(220,30)
\Vertex(180,30)2 
\Photon(180,30)(180,70)2 4

\Line(235,30)(245,30)
\Line(240,34)(240,26)

\Line(260,30)(300,30)
\Line(300,30)(340,30)
\Vertex(300,30)2 
\Photon(300,50)(300,70)2 2
\Line(301,30)(301,50)
\Line(299,30)(299,50)
\Vertex(300,50)2 

\put(170,10){(e)}

\end{picture}
\end{center}
\caption[Compton scattering diagrams for the meson resonances.]
{\label{fig:CSFeynDiag}
\sf Compton scattering diagrams for the meson resonances. 
(a) Elastic diagram.  (b) Pseudoscalar seagull
diagram. (c) Vector resonance seagull diagram. (d) Axial-vector
resonance intermediate state diagram. (e) Pion form factor diagrams.}
\end{figure}
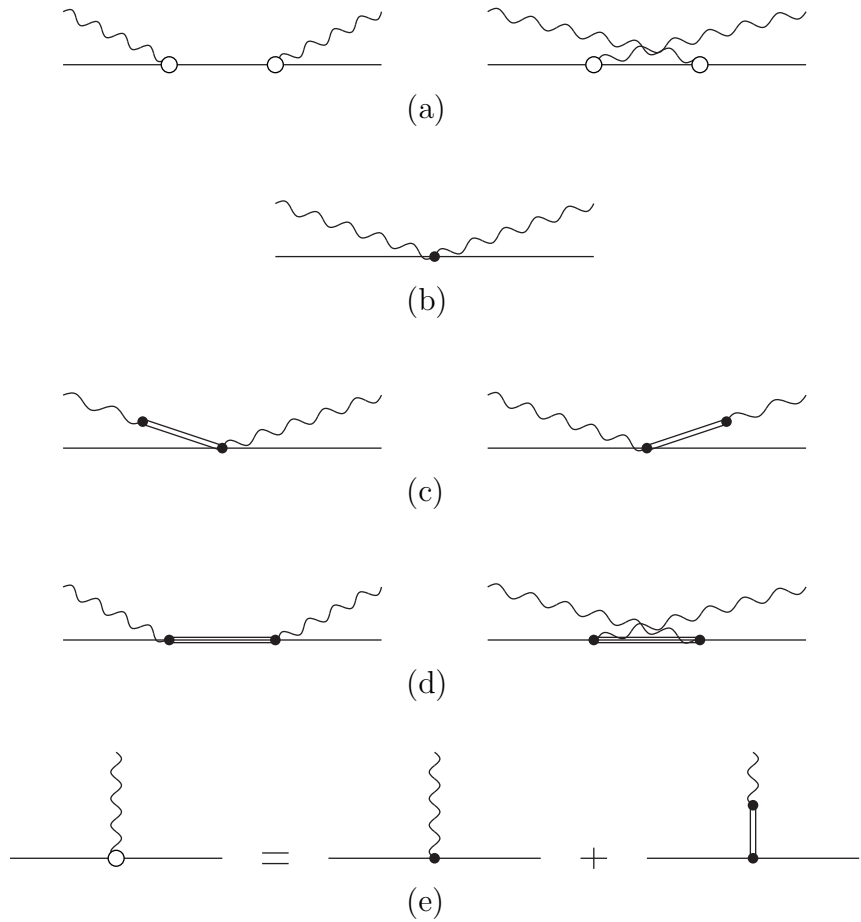

The Feynman diagrams which contribute to the pion Compton scattering
amplitude, given by the above Lagrangian, are shown in
Fig.~\ref{fig:CSFeynDiag}. It is convenient to classify these
diagrams in three groups, which correspond with three gauge invariant
terms of the amplitude.

The first term encloses the contribution given by
Fig.~\ref{fig:CSFeynDiag}.a, and Fig.~\ref{fig:CSFeynDiag}.b,

{\everymath{\displaystyle}
\beq
\begin{array}{l}
T_{\mu\nu}^{(1)}(q^2,\pq) \ \ = \ \ - \: 2 \: D_{1_{\mu\nu}} 
\\ \\  \qquad
  + \: 4 m_\pi^2 |G_\pi(q^2)|^2 
  \left(  {{1}\over{m_\pi^2 - (p+q)^2 -i \epsilon}} 
  + {{1}\over{m_\pi^2 - (p-q)^2 -i \epsilon}} \right) 
   D_{2_{\mu\nu}} \ \ ,
\end{array}
\label{CSA1}
\eeq
}

\noindent
where we have used the vector resonance dominance approximation in the
pion form factor, i.e.,

\beq 
{{F_V G_V}\over{F_{\pi}^2}} = 1 \ \ .
\label{VMD}
\eeq

\ni
This is equivalent to saturating the pion form factor with the rho meson
resonance,

\beq
G_\pi(q^2) = {{m_\rho^2}\over{m_\rho^2 - q^2}} \ \ .
\eeq

The second term, which we call the vector
seagull, is given in Fig.~\ref{fig:CSFeynDiag}.c. It differs from the
pseudoscalar seagull because one of the photon lines interacts through a vector
resonance,

\beq
T_{\mu\nu}^{(2)}(q^2,\pq) = 
  - \: 2 \: {{F_V^2}\over{F_\pi^2}} {{q^2}\over{m_\rho^2 - q^2}} 
  \: D_{1_{\mu\nu}} \ \ .
\label{CSA2}
\eeq

\ni
The third group, due to the axial-vector intermediate state, given by
the diagrams in Fig.~\ref{fig:CSFeynDiag}.d is

{\everymath{\displaystyle}
\beq
\begin{array}{l}
T^{(3)}_{\mu \nu}(q^2,\pq) =  
\\ \\ 
  \qquad
  {{F_A^2}\over{F_\pi^2 m_A^2}}
  \left( { {\left( \pq + q^2 \right)^2 
  +q^2 \left( m_A^2 -(p+q)^2\right)}\over{m_A^2 -(p+q)^2 -i \epsilon}}
  \right.
  \\ \\
  \qquad
  \left. 
   \qquad  + \qquad \quad {{\left( \pq - q^2 \right)^2 
  + q^2 \left( m_A^2 -(p-q)^2\right)}\over{m_A^2 -(p-q)^2 -i \epsilon}} 
  \right) {D_1}_{\mu \nu} 
\\ \\
\qquad
  + {{F_A^2}\over{F_\pi^2 m_A^2}}
 \left( {{- m_\pi^2 q^2} \over{m_A^2 - (p+q)^2 -i \epsilon }} 
      + {{- m_\pi^2 q^2} \over{m_A^2 - (p-q)^2 -i \epsilon }}
 \right) {D_2}_{\mu \nu} \ \ .
\end{array}
\label{CSA3}
\eeq
}

\ni
The numerical values that we use for the parameters involved in the
previous five equations are

{\everymath{\displaystyle}
\begin{eqnarray}
m_\pi  & = & m_{\pi^\pm} \ = \ 0.13956995 
  \pm 0.00000035 \ \mbox{GeV} \ \ , \nonumber \\
F_\pi  & = & 0.0924 \pm 0.0003 \ \mbox{GeV} \ \ , \nonumber \\
m_V    & = & 0.7699 \pm 0.0008 \ \mbox{GeV} \ \ \mbox{and} \nonumber \\
F_V    & = & 0.1529 \pm 0.0036 \ \mbox{GeV} \ \ .
\label{piVparam}
\end{eqnarray}
}

\noindent
For the numerical results in the soft-pion limit, we use the
axial-vector resonance parameters, $m_A$, and $F_A$, obtained by the
Weinberg sum rules \cite{Wein:67} in the narrow-width approximation,

{\everymath{\displaystyle}
\begin{eqnarray}
F_A^{(WSR)}  & = & \sqrt{F_V^2 - F_\pi^2} 
\ \ = \ \ 0.1218 \pm 0.0045 \ \mbox{GeV} \ \ ,
\nonumber \\
m_A^{(WSR)}  & = & {{F_V \; m_V}\over{F_A}} 
\ \ = \ \ 0.9664 \pm 0.0427 \ \mbox{GeV} \ \ .
\label{FamaWSRNW}
\end{eqnarray}
}

\ni
The $p^2$ corrections to the final $m_A$, and $F_A$ which we use in our
final numerical answer at the end of this section, even though
necessary to cancel divergences, are minimal and also have a minute
effect on the numerical results for $\Delta m_\pi$.

\bigskip
\bigskip
\ni
{\it Beyond the narrow-width approximation}
\bigskip

The above analysis utilizes zero-width $\rho$ and $a_1$ resonances.
This ``narrow-width'' approximation is a poor description for these
resonances since they are not particularly narrow, especially the
$a_1$.  The full resonance spectrum can be taken into account by
employing the spectral function - K\"{a}llen-Lehmann - representation
\cite{Bart:65,Bjor:65}.  Furthermore, this representation includes the
effect of higher mass resonances with the same quantum numbers such as
$\rho^\prime$'s in the vector case. The spectral function
representation of these resonances generalizes the spin-1 resonance
propagators encountered in the Compton scattering amplitudes,
$T^{(i)}_{\mu \nu}(q^2, p \cdot q)$, (for $i = 1$ to $3$) given above,

\beq
{{1}\over{m^2 - q^2 - i \epsilon}} \rightarrow
\int \! \! ds {{\rho^R(s)}\over{s - q^2 - i \epsilon}} \ \ .
\label{defspfn}
\eeq

\ni
The sum of the three terms, equations
(\ref{CSA1}, \ref{CSA2}, \ref{CSA3}), of the pion forward Compton
scattering amplitude in the spectral function representation reads

{\everymath{\displaystyle}
\begin{eqnarray}
T_{\mu\nu}(q^2,\pq) & = & 
D_{1_{\mu\nu}} \left\{
  -2 - {{2}\over{F_\pi^2}} 
  \int_0^\infty \! \! ds \rho^R_V(s) 
  {{q^2}\over{s - q^2}} 
  \right. 
\nonumber \\
  & & \left. \ + {{1}\over{F_\pi^2}} \int_0^\infty \! \! ds 
  {{\rho^R_A(s)}\over{s}}
  \left( { {\left( \pq + q^2 \right)^2 
  +q^2 \left(s -(p+q)^2\right)}\over{s -(p+q)^2 -i \epsilon }}
  + (q \rightarrow - q)  \right) \right\}
\nonumber \\
& + & D_{2_{\mu\nu}} \left\{
  4 m_\pi^2 |G_\pi(q^2)|^2 
  \left(  {{1}\over{m_\pi^2 - (p+q)^2 -i \epsilon}} 
  + (q \rightarrow - q) \right)
  \right.
\nonumber \\
& & \left. \ + {{1}\over{F_\pi^2}} \int_0^\infty \! \! ds 
  {{\rho^R_A(s)}\over{s}} 
  \left( {{- m_\pi^2 q^2} \over{s - (p+q)^2 -i \epsilon }} 
  + (q \rightarrow - q)  \right) \right\} \ \ .
\label{CSASpFn}
\end{eqnarray}
}

The narrow-width result can be readily reproduced by letting,

\beq
\rho_{V,A}^{R^{(NW)}}(s) = F_{V,A}^2 \delta ( s - m_{V,A}^2 ) \ \ ,
\label{defNWSpFn}
\eeq

\ni
where the values for $F_V, F_A, m_V$ and $m_A$ are given in
Eqs.~(\ref{piVparam}, \ref{FamaWSRNW}).

\bigskip
\bigskip
\ni
{\it The Cottingham approach}
\bigskip

Since we have a complete form for the Compton Scattering amplitude we
could directly calculate the pion EM mass difference with
Eq.~(\ref{dmdirect}), avoiding the Cottingham method altogether.
Nevertheless, the Cottingham method \cite{Cott:63} allows us to gain
control and insight into the calculation. This method requires the
break down of the scattering amplitude into subtraction and structure
functions terms, which are easily extracted from Eq.~(\ref{CSASpFn}),

{\everymath{\displaystyle}
\begin{eqnarray}
T_1(-Q^2,0) & = & -2 + {{2}\over{F_\pi^2}} 
  \int_0^\infty \! \! ds \rho^R_V(s) {{Q^2}\over{s+Q^2}}
\nonumber \\
& & - {{2}\over{F_\pi^2}} \int_0^\infty \! \! ds \rho^R_A(s)
  {{Q^2}\over{s -p^2 +Q^2}} \left( 1 - {{p^2}\over{s}} \right) \ \ ,
\nonumber \\   
W_1(-Q^2,\nu) & = & {{1}\over{F_\pi^2}} \int_0^\infty \! \! ds
{{\rho^R_A(s)}\over{s}} \ ( \nu -Q^2 )^2 \ 
\delta ( s -p^2 +Q^2 - 2 \nu ) \ \ ,
\nonumber \\   
W_2(-Q^2,\nu) & = & 4 m_\pi^2 
  \left( {{m_\rho^2}\over{m_\rho^2 + Q^2}} \right)^2 
  \delta (Q^2 - 2 \nu ) 
\nonumber \\
& & + {{1}\over{F_\pi^2}} \int_0^\infty \! \! ds {{\rho^R_A(s)}\over{s}}
  \ p^2 Q^2 \ 
  \delta (s -p^2 +Q^2 - 2 \nu ) \ \ ,
\label{SpFnTW}
\end{eqnarray}
}

Before we describe each of these terms in detail, it is useful to be
more familiar with their domain in the $(\nu, Q^2)$ plane. The
subtraction term is the value of $T_{1}$ along the negative $Q^2$
axis. The structure functions are limited by kinematics to a sector of
the first quadrant, their domain is better understood if we introduce
the Bjorken scaling variable, $x = {{Q^2}\over{2 \nu }}.$ The
allowed kinematic ranges for the variables involved are:

\beq
0 \le Q^2 \le \infty \ ; 
c \le \nu \le \infty \ ; \ \mbox{and} \ 
0 \le x \le 1 \ , \ \ \mbox{where} \ c = {{Q^2}\over{2}} \ \ .
\eeq

\begin{figure}[htbp]
\centerline{\psfig{figure=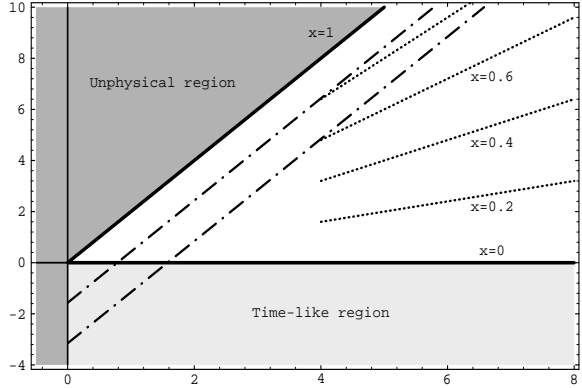,height=2.00in,width=3.00in}}
\caption[]
  {\sf $(\nu,Q^2)$ plane. Units in $GeV^2.$ Unshaded region is the domain
of the structure functions. Solid lines are for $x =$ 0, 0.2, 0.4, 0.6,
0.8 and 1.0. Dashed lines are for $M_R^2 =$ 1.6 and 3.2 $GeV^2.$}
\label{fig:Q2nuplane} 
\end{figure}

The domain of both structure functions in the $(\nu,Q^2)$ plane,
covers the area in the first quadrant which lies between the positive
$\nu$ axis ($x = 0$) and the elastic line, ($x = 1$). This is the
unshaded region shown in Fig.~\ref{fig:Q2nuplane} for pion
kinematics. Within this sector, the figure shows other lines of
constant-$x$ which help describe the structure functions in the
scaling region. It also shows two lines of constant $s = m_R^2$ which
mark the region where the resonant intermediate states are the
dominant contribution.

The scaling region for nucleons is the region above $Q^2 \sim 1$ GeV.
It is described by perturbative QCD. The relevant
degrees of freedom are quarks and gluons, and the structure
functions are described in terms of quark distribution functions,
which depend only on $x$ if we neglect logarithmic deviations. In this
approximation, the structure functions are constant along the
constant-$x$ lines.

The resonance region in the $(\nu,Q^2)$ plane is described by the two
dashed lines parallel to the elastic or $x = 1$ line. These lines
satisfy the equation for constant squared invariant mass of intermediate
resonant states,

\beq
M_R^2 = (p + q)^2 = p^2 + 2 \nu - Q^2 \ \ . 
\eeq

\noindent
Since the graph is for pion values we choose the $a_1$ as the first
resonance, $m_{a_1} = 1.26$ GeV. We also include a second resonance
with mass $m_R = \sqrt{2} \: m_{a_1}$ in order to show the position
of possible higher resonances.

This resonance region is described by chiral Lagrangians which include
spin-1 resonances such as Eq.~(\ref{Lchires}),
\cite{Ecke:89a,Ecke:89b}.  The structure functions obtained through
the chiral Lagrangians in the narrow-width approximation are
constrained to the elastic line, the $a_1$ line, plus other parallel
lines corresponding to possible higher resonances. If we include a
finite width for the resonances, these lines become bands whose
thickness is proportional to the resonance width.

The usefulness of applying the Cottingham method arises from the
breakdown of the scattering amplitude into the three terms shown in
Eq. (\ref{SpFnTW}). We gain control because we can make
reasonable assumptions and establish constraints on the pion
structure functions. At the same time, it is possible to relate the
subtraction term to the soft-pion limit.

All of the resonances couplings will contain form-factors which
suppress the effect of an individual resonance as $Q^2 \to \infty$. We
will assume that the fall-off of all such photon form-factors will
involve a scale which is a typical vector meson mass. We now turn to
the procedure to introduce these form-factors in our dispersive
framework. This has a subtlety in that some naive structures for this
form-factor could upset the soft-pion limits in our formulas. We will
chose a form which is well behaved in the soft-pion limit. The
form-factor also solves what would appear to be a problem in the
present inclusion of resonances, i.e. the structure function
$W_1(Q^2,\nu)$ given in Eq. (\ref{SpFnTW}) has terms proportional to
$\nu^2$ and $Q^4$ which would generate divergences for large $Q^2$.
This is does not occur in the presence of the form-factors. 
This divergent behavior is clear if we calculate
$W_1(Q^2,\nu)$ along the lines of constant $s = m_R^2 \:$. We use the
delta function to eliminate the $\nu$ dependence, i.e., $ \nu =
{{1}\over{2}} (s - p^2 + Q^2) \:$, to obtain
 
\beq
W_1^{(cst \: s)}(Q^2, \nu) = {{1}\over{F_\pi^2}} {{\rho^R_A(s)}\over{s}}
\: {{1}\over{4}} \: (s -p^2 -Q^2)^2 \ \ ,    
\eeq

\noindent
which diverges as $Q^4$. A structure function for a given resonant
state cannot diverge for large $Q^2$ without violating unitarity. For
a single resonance, as it is the case in question, the structure
function must go to zero if we follow a line of constant invariant
mass, $s$, to high energies. In order to achieve this behavior, we
introduce a multiplicative factor which forces its convergence.  This
factor, $K(\nu,s)$ resembles the form factor obtained for the elastic
term through the vector meson dominance model. In addition, it has the
following properties,

{\everymath{\displaystyle} 
\begin{eqnarray}
K(\nu = 0, s) & = & 1 \ \ ,
\nonumber \\
\lim_{Q^2 \rightarrow \infty} K(\nu,s)^{(cst. \: s)} 
 & \sim & {{1}\over{Q^6}} \ \ ,
\nonumber \\
\left[ K(\nu,s)^{(cst. \: s)} \right]_{Q^2=0} & = & 1 \ \ .
\end{eqnarray}               
}                             

\begin{figure}[htbp]
\centerline{\psfig{figure=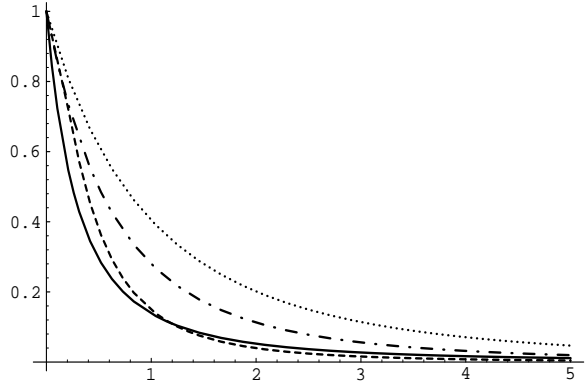,height=2.00in,width=3.00in}}
\caption[]
{\sf Factor $K(\nu,s)$ and pion EM form factor.
Pion EM form factor, solid line. $K(\nu,m_\pi^2),$ dashed line.
$K(\nu,1 \ \mbox{GeV}^2)$, dash-dotted line. $K(\nu,2 \
\mbox{GeV}^2)$, dotted line. Horizontal scale represents $Q^2$ in
$\mbox{GeV}^2.$}
\label{fig:Kfactor} 
\end{figure}

These properties ensure that the subtraction term is left
unchanged, and that the structure function will converge for large
$Q^2.$ The form factor is normalized in order to agree with the
previous result at $Q^2 = 0$ for fixed s. The form factor that
satisfies these conditions is 

\beq
K(\nu,s) = \left( {{m_V^2}\over{m_V^2 + 2 \nu}} \right)^4
 \left( 1  + \eta {{2 \nu}\over{s}} \right) \ \ ,
\label{defKfactr}
\eeq

\noindent
where

\beq
\eta =  {{s}\over{s - p^2}} 
  \left[ \left( 1 + {{s - p^2}\over{m_V^2}} 
  \right)^4 -1 \right] \ \ ,
\eeq

\noindent
and $p^2 = m_\pi^2 \, .$ We have chosen the appropriate value for the
vector meson mass, $ m_V = m_\rho \, .$ The factor $K(\nu,s)$ above,
also has the property of being very close to the $\rho$ contribution to
the pion EM form factor for $s = p^2,$ as it can be seen in
Fig.~\ref{fig:Kfactor}.

The inclusion of this factor in our analysis is easily achieved
through the substitution

\beq
\rho^R_A(s) \to \rho^R_A(s) \: K(\nu,s) \ \ .
\eeq

\ni
The structure functions and subtraction terms read

{\everymath{\displaystyle}
\begin{eqnarray}
T_1(-Q^2,0) & = & -2 + {{2}\over{F_\pi^2}} 
  \int_0^\infty \! \! ds \rho^R_V(s) {{Q^2}\over{s +Q^2}} 
\nonumber \\
& & - {{2}\over{F_\pi^2}} \int_0^\infty \! \! ds \rho^R_A(s)
  {{Q^2}\over{s -p^2 +Q^2}} 
  \left( 1 - {{p^2}\over{s}} \right) \ \ ,
\nonumber \\   
W_1(-Q^2,\nu) & = & {{1}\over{F_\pi^2}} \int_0^\infty \! \! ds
{{\rho^R_A(s)}\over{s}} \: K(\nu,s) \:
( \nu -Q^2 )^2 \: \delta ( s -p^2 +Q^2 - 2 \nu ) \ \ ,
\nonumber \\   
W_2(-Q^2,\nu) & = & 4 m_\pi^2 
  \left( {{m_\rho^2}\over{m_\rho^2 + Q^2}} \right)^2 
  \delta (Q^2 - 2 \nu )
\nonumber \\
& & + {{1}\over{F_\pi^2}} \int_0^\infty \! \! ds 
  {{\rho^R_A(s)}\over{s}} \: K(\nu,s) \: p^2 Q^2 \:
  \delta (s -p^2 +Q^2 - 2 \nu ) \ \ .
\label{SpFnTWFF}
\end{eqnarray}
}

The pion EM mass difference for the above functions is readily
obtained with Eq.~(\ref{dmcott}). We choose to break it into
terms corresponding to those shown in the above equation with an extra
subdivision of the $W_2$ contribution which isolates the elastic term

{\everymath{\displaystyle}
\begin{eqnarray}
\Delta m_{\pi}^2 (Subtr.) & = & 
  {{\alpha}\over{4 \pi}} \int_0^\infty \! dQ^2
  {{3}\over{F_\pi^2}} \left[ F_\pi^2 - \int_0^\infty \! \! ds \rho^R_V(s)
  {{Q^2}\over{s +Q^2}} \right.
\nonumber \\
& & \left. \qquad \qquad
  \ + \int_0^\infty \! \! ds \rho^R_A(s) {{Q^2}\over{s -p^2 +Q^2}} 
  \left( 1 - {{p^2}\over{s}} \right) \right] \ \ ,
\nonumber \\   
\Delta m_{\pi}^2 (W_1) & = & 
  {{\alpha}\over{4 \pi}} \int_0^\infty \! dQ^2
  {{6}\over{F_\pi^2}} \int_0^\infty \! \! ds {{\rho^R_A(s)}\over{s}}
  {{1}\over{\Delta^2}} \left( {{\Delta^2}\over{2}} -Q^2 \right)^2
  \Lambda_1 \! \left( {{\Delta^4}\over{4 p^2 Q^2}} \right) 
  K \left({{\Delta^2}\over{2}},s \right) \ \ ,
\nonumber \\   
\Delta m_{\pi}^2 (Elast.) & = & 
  {{\alpha}\over{4 \pi}} \int_0^\infty \! dQ^2
  \left( {{m_\rho^2}\over{m_\rho^2 + Q^2}} \right)^2 \
  \Lambda_2 \! \left( {{Q^2}\over{4 p^2}} \right) \ \ ,
\nonumber \\
\Delta m_{\pi}^2 (W_2) & = & 
  {{\alpha}\over{4 \pi}} \int_0^\infty \! dQ^2
  {{1}\over{2 F_\pi^2}} \int_0^\infty \! \! ds {{\rho^R_A(s)}\over{s}}
  \: \Delta^2 \: 
  \Lambda_2 \! \left( {{\Delta^4}\over{4 p^2 Q^2}} \right) 
   K \left( {{\Delta^2}\over{2}},s \right) \ \ ,
\label{SpFndmpi}
\end{eqnarray}
}

\noindent
where $\Delta^2 = s - p^2 + Q^2,$ and $\Lambda_i(y)$ (for $i = 1,2 \ $)
are defined in Eq.~(\ref{defL1L2}).

\bigskip
\bigskip
\ni
{\it High energy constraints}
\bigskip

Even though, we can obtain an explicit formula for the pion EM mass
difference by adding all the contributions in Eqs.~(\ref{SpFndmpi}),
it is necessary to analyze the upper limit of the $Q^2$ integral.
Adding all the different contributions in Eqs.~(\ref{SpFndmpi}), and
expanding the $Q^2$ integrand in powers of $1/Q^{2n}$, we obtain,

{\everymath{\displaystyle}
\begin{eqnarray}
& - & \left(  {{3 \alpha }\over{4 \pi F_\pi^2}} \right)
\left[
 - F_\pi^2  + 
 \int_0^\infty \! \! ds \left[ \rho^R_V(s) - 
   \rho^R_A(s)  \left(1 - {{p^2}\over{s}} \right) \right] 
\right] 
\ {{1}\over{Q^0}}
\nonumber \\
& + & \left(  {{3 \alpha }\over{4 \pi F_\pi^2}} \right)
\int_0^\infty \! \! ds \, s \left[ \rho^R_V(s) - 
  \rho^R_A(s)  \left(1 - {{p^2}\over{s}} \right)^2 \right]
\  {{1}\over{Q^2}} \ + \ {\cal O} \left( {{1}\over{Q^4}} \right) \ \ ,
\label{dmpiNWQ2exp}
\end{eqnarray}
}

\ni
where, $p^2 = m_\pi^2$. If the first two terms are not zero, they
originate linear and logarithmic divergences respectively. In order to
obtain a finite pion EM mass difference we cancel them explicitly
generating two constraint equations,

{\everymath{\displaystyle}    
\begin{eqnarray}              
\int_0^\infty \! \! ds \left[ \rho^R_V(s) - 
  \rho^R_A(s)  \left(1 - {{p^2}\over{s}} \right) \right] 
  & = & F_\pi^2 \ \ ,
\label{SpFnnuConstFF}  \\
\int_0^\infty \! \! ds \, s \left[ \rho^R_V(s) - 
  \rho^R_A(s)  \left(1 - 2 {{p^2}\over{s}} + 
  {{p^4}\over{s^2}} \right) \right] & = & 0 \ \ .
\label{SpFnQ2ConstFF} 
\end{eqnarray}
}

\ni
These high $Q^2$ constraints become the Weinberg sum rules in the soft
pion limit, which in the above equations is obtained by letting $p^2 =
0$. We will see later a more detailed explanation of this limit and
its relation to the subtraction term.

Due to the introduction of the convergence factor $K( \nu, s)$, the
above divergences originate only from the subtraction term. We
incorporate the high $Q^2$ constraints in the subtraction term of the
pion EM mass difference Eq.~(\ref{SpFndmpi}) in order to make all
the contributions finite. The following procedure removes both
divergences.

Subtract the linear divergence from the subtraction term by means of
Eq.~(\ref{SpFnnuConstFF}),

{\everymath{\displaystyle}
\begin{eqnarray}
\Delta m_{\pi}^2 (Subtr.) & = & 
{{\alpha}\over{4 \pi}} \int_0^\infty d Q^2 {{3}\over{F_\pi^2}} \left\{
F_\pi^2 - \int_0^\infty \! \! ds \rho^R_V(s) {{Q^2}\over{s +Q^2}} 
\right.
\nonumber \\
& & + \int_0^\infty \! \! ds \rho^R_A(s) {{Q^2}\over{s -p^2 +Q^2}} 
  \left( 1 - {{p^2}\over{s}} \right)
\nonumber \\   
& & \left. - \left[ F_\pi^2 - \int_0^\infty \! \! ds \rho^R_V(s) 
  + \int_0^\infty \! \! ds \rho^R_A(s) 
  \left( 1 - {{p^2}\over{s}} \right) \right] \right\}
\nonumber \\   
& = & {{\alpha}\over{4 \pi}} \int_0^\infty d Q^2 {{3}\over{F_\pi^2}}
  \left\{ \int_0^\infty \! \! ds \rho^R_V(s) {{s}\over{s +Q^2}} 
  \right.
\nonumber \\
& & \left. - \int_0^\infty \! \! ds \rho^R_A(s) 
  {{s - p^2}\over{s -p^2 +Q^2}} \left( 1 - {{p^2}\over{s}}
  \right) \right\} \ \ .
\label{dmpiSublindiv}
\end{eqnarray}
}

\noindent
Integrate over $Q^2$ and cancel the correspondent logarithmic
divergence by subtracting Eq.~(\ref{SpFnQ2ConstFF}) multiplied by
ln $\Lambda_{Q^2}$ ,

{\everymath{\displaystyle}
\beq
\begin{array}{l}
\Delta m_{\pi}^2 (Subtr.)  =  
\\ \\
  \qquad  
  \lim_{\Lambda_{Q^2} \rightarrow \infty}
  {{\alpha}\over{4 \pi}} {{3}\over{F_\pi^2}} \left\{ \int_0^\infty
  \! \! ds s \rho^R_V(s) \ln {{\Lambda_{Q^2} + s}\over{s}} 
  \right.
  \\ \\
  \qquad
  - \int_0^\infty \! \! ds \rho^R_A(s) (s -p^2 ) 
  \left( 1 - {{p^2}\over{s}} \right) 
  \ln {{\Lambda_{Q^2} + s - p^2}\over{s -p^2}} 
\\ \\ 
  \qquad 
  - \left. \left[ \int_0^\infty \! \! ds s \rho^R_V(s) \ln \Lambda_{Q^2} 
  - \int_0^\infty \! \! ds \rho^R_A(s) (s -p^2)^2 
  \ln \Lambda_{Q^2} \right] \right\} \ \ .
\end{array}
\eeq
}

\noindent
Add the terms and take the limit $\Lambda_{Q^2} \rightarrow 
\infty,$ to obtain 

{\everymath{\displaystyle}
\beq
\begin{array}{l}
\Delta m_{\pi}^2 (Subtr.) = 
\\
\\ 
 \qquad
  - {{\alpha}\over{4 \pi}} {{3}\over{F_\pi^2}} 
  \int_0^\infty \! \! ds \left[ s \ln s \ \rho^R_V(s) 
  - (s -p^2 ) \ln (s - p^2) \rho^R_A(s) 
  \left( 1 - {{p^2}\over{s}} \right) \right] \ \ .
\end{array}
\label{SubtSpFndmpi} 
\eeq
}

\noindent
The above contribution is free of divergences. Furthermore, in the
soft-pion limit, i.e. $p^2=0$, it is equivalent to the Das et al.
calculation \cite{Das:67}. Finally, we have a useful formula to
calculate $\delta m_\pi^{(EM)}$ free of divergences, which was mainly
the product of the Lagrangian introducing the chiral couplings of the
spin-1 resonances. We proceed to show the close relation of the
subtraction term and the soft-pion limit and to see how we reproduce
the results of ${\cal O}(p^4)$ chiral perturbation theory with the
above scattering amplitude.

\bigskip
\bigskip
\ni
{\it Soft-pion limit and its relation to the subtraction term}
\bigskip

In the following we will show that the subtraction term is given by
the soft-pion limit up to corrections of order $p^2$. In this
discussion we refer to non-contact contributions as all contributions
except the pion seagull term. This term is the only one which has both
photons interacting at the same vertex and therefore we treat it
differently in the following discussion.

The non-contact contributions to the Compton scattering amplitude have
the form

\beq
T^{(NC)}_{\mu\nu}(q^2,\pq) = i \int \! \! dx e^{-iqx}
\langle \pi | T \left( V_\mu(x) V_\nu(0) \right) | \pi \rangle \ \ .
\label{defpiCSAInt}
\eeq

\noindent
Consider the soft-pion theorem \cite{Adle:68},

\beq
\lim_{p_\mu \rightarrow 0} \langle \pi^k(p) \beta | O | \alpha \rangle
  = - {{i}\over{F_\pi}} \langle \beta | 
  \left[ Q_5^k, O \right] | \alpha \rangle \ \ ,
\label{SftPiThm}
\eeq

\noindent 
where $\beta$ and $\alpha$ are arbitrary states and $Q_5^k = \int \!
\! d^3x  A_0^k(x)$ is an axial charge. We also need the commutators

\beq
\left[ Q_5^i, V_\mu^j \right] \ \ = \ \ i f^{ijk} V_\mu^k \ \ , 
\qquad
\left[ Q_5^i, A_\mu^j \right] \ \ = \ \ i f^{ijk} A_\mu^k \ \ .
\eeq

\noindent
The result of applying the soft-pion theorem to
Eq.~(\ref{defpiCSAInt}) is 

\beq
\lim_{p_\mu \rightarrow 0} T^{(NC)}_{\mu\nu}(q^2,\pq) 
= - i \int \! \! d^4 x e^{iqx} {{- 2}\over{F_\pi^2}} 
  \langle 0 | 
  T \left( V_3^\mu(x) V_3^\nu(0) - A_3^\mu(x) A_3^\nu(0) \right)
  | 0 \rangle \ \ .
\label{sftpider1}
\eeq

\noindent
The two-current time ordered products are related to the spectral
functions by

{\everymath{\displaystyle}
\begin{eqnarray}
\langle 0 | T \left( V_a^\mu(x) V_b^\nu(0) \right) | 0 \rangle
  & = & i \delta_{ab} \int_0^\infty \! \! ds \rho_V(s) 
 \left( \Box g_{\mu\nu} - \partial^\mu \partial^\nu \right)
 \int {{d^4 k}\over{(2 \pi)^4}} {{e^{-ikx}}\over{k^2 -s +i \epsilon}}
 \ \ , 
\nonumber \\
\langle 0 | T \left( A_a^\mu(x) A_b^\nu(0) \right) | 0 \rangle
 & =  & - i \delta_{ab} F_\pi^2 \partial^\mu \partial^\nu
 \int {{d^4 k}\over{(2 \pi)^4}} {{e^{-ikx}}\over{k^2 +i \epsilon}}
\nonumber \\
 & & \! \! \! \! \! \! \! \! \! \! \! \! \! \! \! \! \! \! \! 
  \! \! \! \! \! \! \! \! \! \! \! \! \! \! \! \! \! \! \! \!
  \! \! \! \! \! \! \! \! \! \! \! 
+ i \delta_{ab} \int_0^\infty \! \! ds \rho_A(s) 
 \left( \Box g_{\mu\nu} - \partial^\mu \partial^\nu \right)
 \int {{d^4 k}\over{(2 \pi)^4}} {{e^{-ikx}}\over{k^2 -s +i \epsilon}}
\ \ .
\label{twocurr}
\end{eqnarray}
}

\ni
Upon combining Eqs.~(\ref{sftpider1}) and (\ref{twocurr}),
integrating over $d^4 x,$ and using the resulting $\delta$ function
to integrate over $d^4 k$, we obtain 

{\everymath{\displaystyle}
\beq
\begin{array}{l}
\lim_{p_\mu \rightarrow 0} T^{(NC)}_{\mu\nu}(q^2,\pq) \ \ =
\\ \\ \qquad
  - 2 {{q^\mu q^\nu}\over{q^2 + i \epsilon}} 
  + {{2}\over{F_\pi^2}} \int_0^\infty \! \! ds 
  \left( -g^{\mu\nu} + {{q^\mu q^\nu}\over{q^2}} \right)
  \left( \rho_V(s) - \rho_A(s) \right)
  {{q^2}\over{q^2 -s +i \epsilon}} \ \ .
\end{array}
\eeq
}

\ni
The contact term or pion seagull contribution to the pion Compton
scattering amplitude is

\beq
T^{(C)}_{\mu\nu}(q^2,\pq) = 2 g_{\mu\nu} \ \ ,
\label{defpiCSACont}
\eeq

\noindent
which remains unchanged in the soft-pion limit. Adding both
contributions to the pion Compton scattering amplitude in the
soft-pion limit, we obtain

\beq
\lim_{p_\mu \rightarrow 0} T_{\mu\nu}(q^2,\pq) 
  = D_{1_{\mu\nu}} \left\{ - 2 
  + {{2}\over{F_\pi^2}} \int_0^\infty \! \! ds 
  ( \rho_V(s) - \rho_A(s) )  {{q^2}\over{q^2 -s}} \right\} \ \ .
\label{CSASftPi}
\eeq

An alternative way of reproducing the soft-pion limit
result above, is letting $p_\mu \rightarrow 0$ in Eq.~(\ref{CSASpFn}). In
order to implement this limit the following relations are useful,

{\everymath{\displaystyle}
\begin{eqnarray}
\lim_{p_\mu \rightarrow 0} D_{1_{\mu\nu}} & = & D_{1_{\mu\nu}} \ \ , 
\nonumber \\
\lim_{p_\mu \rightarrow 0} p^2 D_{2_{\mu\nu}} & = & 0 \ \ , 
\nonumber \\
\lim_{p_\mu \rightarrow 0} T_1(q^2,\pq) 
& = & \left. T_1(q^2,0)  \right|_{p^2 = 0}  \ \ . 
\label{SftPiRel}
\end{eqnarray}
}

\noindent
From these relations it follows that the only surviving term in this
limit is the subtraction term $T_1(q^2,0),$

\beq
\lim_{p_\mu \rightarrow 0} T_{\mu\nu}(q^2,\pq) = 
D_{1_{\mu\nu}} \lim_{p_\mu \rightarrow 0} T_1(q^2,0) \ \ .
\eeq

\noindent
This gives the same result as Eq.~(\ref{CSASftPi}) when we identify

\beq
\lim_{p_\mu \rightarrow 0} 
\left( \rho_V^R(s) - \rho_A^R(s) \right) =
\left( \rho_V(s) - \rho_A(s) \right) \ \ .
\eeq

We can now calculate the soft-pion limit to the pion EM mass difference,

\beq
\lim_{p_\mu \rightarrow 0} \Delta m_\pi^2 = {{\alpha}\over{4 \pi}} 
  \int_0^\infty \! \! dQ^2 {{3}\over{F_\pi^2}} 
  \left\{ F_\pi^2 - \int_0^\infty \! \! ds 
  \left( \rho_V(s) - \rho_A(s) \right)
  {{Q^2}\over{Q^2 + s}} \right\} \ \ ,
\label{dmSubSfPi}
\eeq

\noindent
where $Q^2 = - q^2 $. We follow the procedure described earlier in
order to cancel the linear and logarithmic divergences occurring in
the above equation.  The cancellation of these divergences imposed by
the finiteness of the pion EM mass difference requires

{\everymath{\displaystyle}
\begin{eqnarray}
\int_0^\infty \! \! ds \left( \rho_V(s) - \rho_A(s) \right) 
  & = & F_\pi^2 \ \ ,
\label{W1}
\\
\int_0^\infty \! \! ds s \left( \rho_V(s) - \rho_A(s) \right) 
  & = & 0 \ \ .
\label{W2}
\end{eqnarray}
}

\noindent
These are Weinberg sum rules \cite{Wein:67}, obtained in our case
as a consequence of the finiteness of $\delta m_\pi^{(EM)}$ in the
soft-pion limit. Subtracting the linear and logarithmic divergences in
the same way as for the subtraction term, Eqs.
(\ref{dmpiSublindiv}-\ref{SubtSpFndmpi}), we obtain

\beq
\lim_{p_\mu \rightarrow 0} \Delta m_\pi^2 = 
  - {{\alpha}\over{4 \pi}} {{3}\over{F_\pi^2}} 
  \int_0^\infty \! \! ds 
   s \ln s \left( \rho_V(s) - \rho_A(s) \right) \ \ .
\label{dmpiSftPi}
\eeq

\noindent
This is the result obtained by Das et al. \cite{Das:67}.

Finally, we evaluate Eq.~(\ref{dmpiSftPi}) in the narrow-width
approximation to obtain the numerical result

\beq
\lim_{p_\mu \rightarrow 0} \Delta m_\pi^{(NW)} = 
  4.685 \ \mbox{MeV} \ \ .
\label{dmpiSftPiNW}
\eeq

\bigskip
\bigskip
\ni
{\it Spectral functions}
\bigskip

We seek an improved description of the physics of the resonance region
with the spectral functions $\rho^R_V(s)$ and $\rho^R_A(s)$ replacing
the narrow-width description. The ingredients to the spectral
functions clearly are the same resonance states that are revealed by
the usual vector and axial-vector spectral functions $\rho_V(s)$ and
$\rho_A(s)$. In addition we have just seen that in the soft-pion limit
there is an exact correspondence $\rho^R_V(s) -
\rho^R_A(s) = \rho_V(s) - \rho_A(s)$. This leads us to utilize the
experimental spectral functions determined in reference
\cite{Dono:94a} in order to produce a shape for $\rho^R_V(s)$ and
$\rho^R_A(s)$. In both the soft-pion limit and the full Cottingham
calculation at order $m_\pi^2$, the high energy continuum cancels in
the mass shift. We therefore separate each spectral function into two
contributions, one due to the resonances and the other due to the high
energy continuum common to both vector and axial-vector channels. The
resonant part is chosen to match the resonances revealed in the
phenomenological analysis of the data in \cite{Dono:94a}. These
spectral functions are then slightly altered to obey the full
constraint equations including $p^2$ terms of Eqs. (\ref{RnuConstr})
and (\ref{RQ2Constr}). A continuum contribution, common to both vector
and axial-vector channels, was included in \cite{Dono:94a}, but is
here kept separate from the resonances. The result of this is that we
identify

\beq
\rho_{V,A}(s) = \rho_{V,A}^R(s) + \rho_{V,A}^C(s) \ \ , 
\label{defRhoRC}
\eeq

\noindent
with a continuum contribution
\beq
\rho^C(s) = \rho_V^C(s) = \rho_A^C(s) \ \ .
\eeq

The precise identification of the continuum is not unique, but since
the difference of spectral functions enters, reasonable variations do
not produce a large final effect. The specific form that we use is
shown in Fig. 5.

\begin{figure}[htbp]
\centerline{\psfig{figure=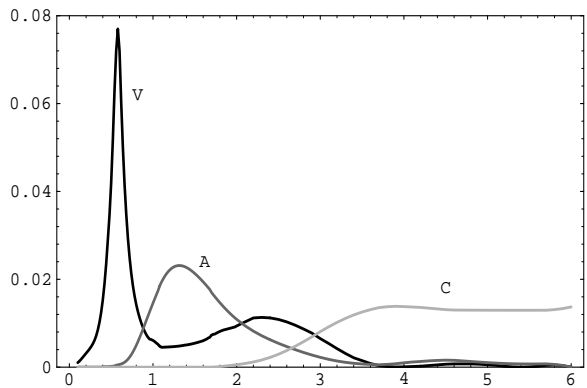,height=2.00in,width=3.00in}}
\caption[]{\sf Vector and axial-vector spectral functions. The graph shows
  (V) $\rho_V^R(s),$ (A) $\rho_A^R(s),$ and (C) $\rho_C(s)$ versus 
  $s$. The $s$ scale is given in $\mbox{GeV}^2.$}
\label{fig:RhoARVRC} 
\end{figure}

It is clear that the greatest source of model dependence in our
calculation comes form the numerical identification described above.
Our procedure in setting up the calculation in the Cottingham method
is very general. However, we do not have directly available the
experimental structure functions for photons scattering off of pions.
We have used an identification which is valid in the chiral limit in
order to provide this numerical input. There could be shifts in the
couplings of these resonances which are of order $m_\pi^2$. These
could provide changes in the final answer at order $m_\pi^2$ which
would be of interest to us. This is partially relieved by the fact
that the analysis of \cite{Dono:94a} was carried out with real world
data, not strictly in the chiral limit. Thus the masses, widths and
shapes of the resonances will accurately reflect physics with $m_\pi^2
\neq 0$. Likewise we know that in the narrow-width approximation we have
the right description, so that we don't see a source of major
uncertainty due to the non-zero widths. This means that our model
dependence comes from possible $m_\pi^2$ dependences in the resonance
couplings, and our implicit assumption is that these are smaller than
the $m_\pi^2$ dependence from the propagators. We have not been able
to find a way to do better than this in the phenomenological analysis.

\bigskip
\bigskip
\ni
{\it Comparison $O(E^4)$ Chiral Perturbation Theory }
\bigskip

We also like to compare our method to the standard chiral perturbation
approach. The lowest energy region of the pion structure function can
be described by the chiral SU(3) Lagrangian to order $p^4$, originally
developed by Gasser and Leutwyler \cite{Gass:84,Gass:85a}. Besides the
elastic and seagull terms, and ignoring pion loops, the only relevant
terms involved in the pion Compton scattering amplitude are the $L_9$
and $L_{10}$ terms,

{\everymath{\displaystyle}
\begin{eqnarray}
{\cal L}_4  =   
  & - & i L_9 \Tr \left( F^{\mu\nu}_R D_\mu U D_\nu U^\dagger
      + F^{\mu\nu}_L D_\mu U^\dagger D_\nu U \right)
\nonumber \\
  & + &  L_{10} \Tr \left( U^\dagger F^{\mu\nu}_R U F_{L\mu\nu} \right)
    + \mbox {\ other} \ .
\label{L9L10}
\end{eqnarray}
}

The pion forward Compton scattering amplitude resulting from this
Lagrangian was calculated by Bijnens and Cornet \cite{Bijn:88}, and
Donoghue et al. \cite{Dono:88}. Their result, up to pion
loop contributions which are small, is

{\everymath{\displaystyle}
\begin{eqnarray}
T_{\mu\nu}(p,q) & = & 
  - {{8 p^2 q^2}\over{q^4 - (2 \pq)^2}} 
    \left( 1 + {{2 L_9^r q^2}\over{F_{\pi}^2}} \right)^2 D_{2_{\mu\nu}}
\nonumber \\
& &  - 2 D_{1_{\mu\nu}}
  + {{8 L_{10}^r q^2}\over{F_{\pi}^2}} D_{1_{\mu\nu}}
  + \mbox{\ loops} \ \  ,
\label{D1D2p4Compt}
\end{eqnarray}
}

\ni
Expanding our narrow-width result in powers of external momenta $p_\mu$ and
$q_\mu$, we obtain,  

{\everymath{\displaystyle}
\beq
\begin{array}{l}
T_{\mu\nu}^{(q^2 \ exp.)}(q^2,\pq) \ = \
  D_{1_{\mu\nu}} \left\{
    -2 -2 {{F_V^2}\over{F_\pi^2}} {{q^2}\over{m_\rho^2}} 
    + 2 {{F_A^2}\over{F_\pi^2}}
    {{q^2}\over{m_A^2}} \right\}
\\ \\
\qquad
  + \ D_{2_{\mu\nu}} \left\{
    4 m_\pi^2 \left( 1 + {{q^2}\over{m_V^2}} \right)^2
    {{- 2 q^2}\over{q^4 - (2 \pq)^2}} 
    \ + 2 {{F_A^2}\over{F_\pi^2 m_A^2}}
    {{- m_\pi^2 q^2} \over{m_A^2}} \right\} 
\\ \\
\qquad
  + \ \mbox{higher order terms in } (q^2,\pq,p^2) \ \ .
\end{array}
\label{CSAexpq2}
\eeq
}

The relations for the $L_{9}$ and $L_{10}$ in terms of the spin-1 resonance
parameters are obtained by inspection from
Eqs.~(\ref{D1D2p4Compt}) and (\ref{CSAexpq2}).

{\everymath{\displaystyle}
\begin{eqnarray}
L_{9}  & = & {{F_\pi^2}\over{2 m_V^2}} \ \ , 
\nonumber \\
L_{10} & = & - {{1}\over{4}} 
  \left( {{F_V^2}\over{m_V^2}} - {{F_A^2}\over{2 m_A^2}} \right) \ \ .
\label{L9L10NW}
\end{eqnarray}
}

\noindent
This result is in agreement with Ecker et al. \cite{Ecke:89a}. The
above equation for $L_{10}$ is also the narrow-width approximation for
the sum rule (W0) in reference \cite{Dono:94a}

Substituting the narrow-width parameters in Eq.~(\ref{L9L10NW}) we obtain,

{\everymath{\displaystyle}
\begin{eqnarray}
L_{9}  & = & ( 7.20 \pm 0.05 ) \times 10^{-3} \ \ , 
\nonumber \\
L_{10} & = & - ( 5.89 \pm 0.65 ) \times 10^{-3} \ \ .
\label{L9L10NWNum}
\end{eqnarray}
}

\noindent
These are seen to be within reasonable agreement with the experimental
values,

{\everymath{\displaystyle}
\begin{eqnarray}
L_{9}  & = &   (7.1 \pm 0.3) \times 10^{-3} \ \ , 
\nonumber \\
L_{10} & = & - (6.84 \pm 0.3) \times 10^{-3} \ \ .
\label{L9L10Expt}
\end{eqnarray}
}

\noindent
The difference between the values in Eq.~(\ref{L9L10NWNum}) and
(\ref{L9L10Expt}) gives an estimate for the loop contributions which
we neglected in the ${\cal O}(p^4)$ chiral Lagrangian calculation of
the Compton scattering amplitude. Besides the loop corrections, the
difference can also be due to the inaccuracy of the narrow-width
approximation.

\bigskip
\bigskip
\ni
{\it Scaling region}
\bigskip

The low and intermediate energy regions of the structure functions are
described above. To complete the analysis of the structure functions
we need to describe the scaling region at large values of $(\nu,
Q^2)$. The ingredients and general behavior in this region are well
known. The structure functions become largely functions of the Bjorken
scaling variable $x = {Q^2}/{2 \nu}$, with logarithmic $Q^2$
variations predictable by QCD \cite{Bjor:69}. This is easy to build
into the Cottingham analysis \cite{Pere:95}. However there is not a
need to describe the details here since the scaling region cancels in
the difference between the charged and neutral pions masses, to the
order that we are working here.

In the limit that the u and d quark masses are equal, the deep
inelastic structure functions of the neutral and charged pions are
equal. This leads to 

\beq
\Delta m_\pi(Scaling) = 0 \ \ .
\eeq 

\ni
To the extent that the u, d masses are different, the structure
functions may differ. However we are calculating the electromagnetic
effect in the limit $m_u = m_d$, so that we are not sensitive to this
effect.

\bigskip
\bigskip
\ni
{\it $V \rightarrow \pi \gamma$ contribution}
\bigskip

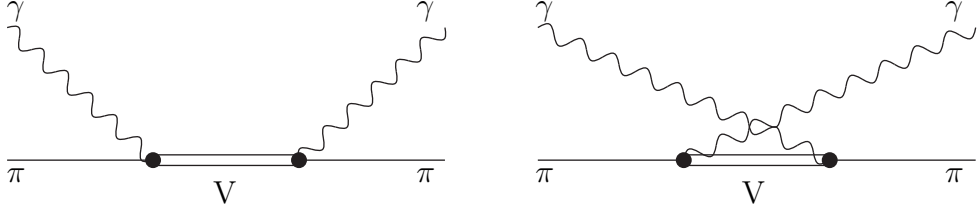
\begin{figure}[htbp]
\ls{1.0}
\begin{center}
\begin{picture}(440,110)(0,0)

\Line(30,50)(85,50)
\Vertex(85,50)3
\Line(85,52)(140,52)
\Line(85,48)(140,48)
\Vertex(140,50)3
\Line(140,50)(195,50)
\Photon(30,100)(85,50)3 6
\Photon(140,50)(195,100)3 6

\Line(230,50)(285,50)
\Vertex(285,50)3
\Line(285,52)(340,52)
\Line(285,48)(340,48)
\Vertex(340,50)3
\Line(340,50)(395,50)
\Photon(230,100)(340,50)3 9
\Photon(285,50)(395,100)3 9
\put(30,42){$\pi$}
\put(185,42){$\pi$}
\put(108,34){V}
\put(30,105){$\gamma$}
\put(185,105){$\gamma$} 
\put(230,42){$\pi$}
\put(385,42){$\pi$}
\put(308,34){V}
\put(230,105){$\gamma$}
\put(385,105){$\gamma$} 
\end{picture}

\end{center}
\caption[Intermediate vector diagrams.]
{\label{fig:intvecdia} 
 \sf Intermediate vector diagrams.
}
\end{figure}

To have a more complete phenomenological description of the Compton
scattering amplitude we also include the effect of intermediate vector
meson diagrams shown in Fig. \ref{fig:intvecdia}. The motivation for
introducing these diagrams is the experimental observation of the
radiative meson decays $\omega \rightarrow \pi \gamma$ and $\rho
\rightarrow \pi \gamma,$ and $\phi \rightarrow \pi \gamma .$ The
effective Lagrangian which includes the $V
\pi \gamma$ vertices is

\beq
{\cal L} = e {{\sqrt{R_V}}\over{2}}
  \epsilon^{\mu \nu \alpha \beta} 
  F_{\mu \nu} V_{\alpha} \partial_\beta \pi \ \ ,
\label{LIntVec}
\eeq 

\noindent
This Lagrangian is invariant under parity and charge conjugation
transformations, as well as under chiral rotations. The choice of
including the EM field strength tensor ensures gauge invariance, and
the pion momentum dependence corresponds to the correct soft-pion
limit for the vertex.

We introduce the spectral functions $g_V(s)$ to describe the
intermediate states in Fig.~\ref{fig:intvecdia}.  The normalization
of these functions is chosen in order to make the subtraction term
contribution compatible with the ones obtained for the axial-vector
case. The narrow-width approximation for $g_V(s)$ is

\beq
g_V(s) \ = \ H_V^2 \delta (s - m_V^2) \ = \ 
           F_\pi^2 R_V  \delta (s - m_V^2) \ \ .
\label{IntVecSpFnNW}
\eeq

\ni
The subtraction term and the structure functions for the intermediate vector
meson diagrams follow from the Lagrangian in Eq.~(\ref{LIntVec}),

{\everymath{\displaystyle}    
\begin{eqnarray}              
T_1 (-Q^2,0) & = & {{2}\over{F_\pi^2}} 
\int_0^\infty \! \! ds g_V(s) {{p^2 Q^2}\over{s - p^2 + Q^2}} \ \ ,
\nonumber \\
W_1 (-Q^2,\nu) & = & {{1}\over{F_\pi^2}} 
\int_0^\infty \! \! ds g_V(s) K(\nu,s)
(\nu^2 + p^2 Q^2) \delta(s - p^2 + Q^2 - 2 \nu) \ ,
\nonumber \\
W_2 (-Q^2,\nu) & = & {{1}\over{F_\pi^2}} 
\int_0^\infty \! \! ds g_V(s) K(\nu,s)
p^2 Q^2 \delta(s - p^2 + Q^2 - 2 \nu) \ ,
\label{IntVecTW}
\end{eqnarray}
}                             

\noindent
where $K(\nu,s)$ is the factor defined in Eq.~(\ref{defKfactr}).  The
factor $K(\nu,s)$ ensures the $Q^2$ convergence of the structure
functions as in the intermediate axial-vector state case.  We only
need to find the spectral function $g_V(s)$ in order to determine the
above functions.

There are four possible vector intermediate states for the pion
Compton amplitude, the $\rho^\pm$ for the charged pions, and the
$\rho^0$, $\omega$ and $\phi$ for the neutral pion. The coupling
constants, $R_{\rho^\pm},$ $R_{\rho^0},$ $R_{\omega},$ and $R_{\phi}$
can be extracted from the radiative decays of these vector mesons. We
refer the reader to references \cite{Dono:93b,Ko:90,ODon:81,Tana:64}
for a review and examples of obtaining such couplings. The couplings
$R_{\rho^\pm}$ and $R_{\rho^0}$ are the same if we take isospin to be
an exact symmetry.  This means that the charged and neutral pion EM
self energies due to the $\rho$ intermediate state would cancel in the
pion EM mass difference. However, the $\omega$ and $\phi$ intermediate
state contributions do not present such a cancellation.  Since
isospin breaking effects are generally of small magnitude, we shall
neglect the intermediate $\rho$ contribution to the pion EM mass
difference.

We can determine the $\omega$ coupling, $R_{\omega}$, from the
experimental measurement of the radiative decay $\omega \rightarrow
\pi^0 \gamma,$

\beq
R_{\omega} \ = \ {{24}\over{\alpha}}
    {{m_\omega^3}\over{(m_\omega^2 - m_\pi^2)^3}} 
    \ \Gamma_{\omega \rightarrow \pi \gamma}
  \ = \ 5.40 \pm  0.32 \ \mbox{GeV}^{-2} \ \ .
\eeq

\ni
Likewise we determine the $\phi$ coupling, $R_{\phi}$,

\beq
R_{\phi} \ = \ {{24}\over{\alpha}}
    {{m_\phi^3}\over{(m_\phi^2 - m_\pi^2)^3}} 
    \ \Gamma_{\phi \rightarrow \pi \gamma}
  \ = \ 0.019 \pm  0.002 \ \mbox{GeV}^{-2} \ \ ,
\eeq

\noindent
where we have used the experimental values listed by the Particle Data Group
\cite{PDG:94}. We do not consider the $\phi$ vector meson intermediate
state further because its coupling is an order of magnitude smaller
than the experimental uncertainty of the $\omega$ coupling.

We are now ready to determine the spectral function $g_V(s).$ Since
the only resonance involved is the $\omega,$ we can safely use the
narrow-width approximation of Eq.~(\ref{IntVecSpFnNW}). The width
of the $\omega$ is only $ 1 \% $ of its mass. This is in contrast with
the $\rho$ and $a_1$ resonances for which the widths are $ 20 \% $ and
$ 33 \% $ of their mass respectively. In the narrow-width
approximation we only need $m_\omega$, taken from \cite{PDG:94}, and 
$H_V$, given by

\beq
H_V  \ = \ F_\pi \sqrt{R_V}
     \ = \ 0.215 \pm 0.013\ \ . 
\eeq

We should be careful when comparing $g_V(s)$ to $\rho_{V,A}(s)$ since
they have different units. The relationship among these structure
functions will become clear in the following subsection.

The intermediate vector meson subtraction term and structure function
contributions to the pion EM mass difference are obtained by combining
Eq.s~(\ref{dmcott}) and (\ref{IntVecTW}),

{\everymath{\displaystyle}
\begin{eqnarray}
\Delta m_{\pi}^2 (Subtr.) & = & 
 {{3 \alpha} \over {4 \pi F_\pi^2}}  
 \int_0^\infty \! \! dQ^2   \int_0^\infty \! \! ds g_V(s) 
 {{p^2 Q^2} \over {s - p^2 + Q^2}} \ \ ,
\nonumber \\
\Delta m_{\pi}^2 (W_1) & = & 
  {{- \alpha}\over{4 \pi}} 
 \int_0^\infty \! \! dQ^2 
  {{6}\over{F_\pi^2}} 
\nonumber \\
  & \times &
  \int_0^\infty \! \! ds g_V(s) 
  K \! \left( {{\Delta^2}\over{2}}, s \right) {{1}\over{\Delta^2}} 
  \left( {{\Delta^4}\over{4}} + p^2 Q^2 \right)
  \Lambda_1 \! \left( {{\Delta^4}\over{4 p^2 Q^2}} \right) \ \ ,
\nonumber \\
\Delta m_{\pi}^2 (W_2) & = & 
  {{- \alpha}\over{4 \pi}} 
  \int_0^\infty \! \! dQ^2 
  {{1}\over{2 F_\pi^2}} \int_0^\infty \! \! ds g_V(s) 
  K \! \left( {{\Delta^2}\over{2}}, s \right) \Delta^2 
  \Lambda_2 \! \left( {{\Delta^4}\over{4 p^2 Q^2}} \right) \ \ ,
\label{dmpiIntVecW12res}
\end{eqnarray}
}

\noindent
where $\Delta^2 = s - p^2 +Q^2$, $p^2 = m_\pi^2$, and the functions
$\Lambda_i(y),$ for $i = 1,2$ are defined in Eq.~(\ref{defL1L2}). The extra
minus sign appears because the vector intermediate state diagrams contribute
to the neutral pion EM self energy.

\bigskip
\bigskip
\ni
{\it General treatment of other possible contributions}
\bigskip

At this point we have a fairly complete calculation of the pion EM
mass difference broken down into different contributions. We have
included the spin-1 resonances through their lowest order chiral
couplings, the scaling, and the intermediate vector resonance
contributions to the pion EM mass difference. By analogy with the
nucleon structure functions, we are comfortable with our estimates of
the structure function contributions. These are small, and even a
factor of two correction would amount to a small correction to the
total mass difference. Therefore, we concentrate in the subtraction
term contribution estimate.

The subtraction term has been obtained by calculating the pion Compton
scattering amplitude with the effective chiral Lagrangian for the vector and
axial-vector resonances of Eq.~(\ref{Lchires}) and the effective Lagrangian
for the intermediate vector meson contribution of Eq.~(\ref{LIntVec}). In
general, there could be other possible contributions to the
subtraction term. These could be introduced by higher order
effective Lagrangians. Their contributions to $\delta m_\pi^{EM}$
would be small since they would be of higher order in the external
momenta $p^2$ and $q^2.$ The terms with higher powers of $p^2$ are
naturally small, otherwise terms of higher order
in $q^2$ are in principle divergent. The finiteness of $\delta
m_\pi^{EM}$ requires that all the higher powers of $q^2$ cancel in the
same way that the order 1 and $1/Q^2$ cancel due to the Weinberg sum
rules in the soft-pion limit case.

We include all other possible contributions to the subtraction term,
not yet accounted for in the previous analysis, by introducing the remainder
term,

\beq
{{2}\over{F_\pi^2}}
\int_0^\infty \! \! ds \: R(Q^2,p^2,s) \ \ .
\eeq

\noindent
The purpose of including this term is to show explicitly the effect of
possible corrections to our current scattering amplitude and its role
in the high energy constraints and final formula for the EM mass
difference.

There are some conditions required upon this remainder term. It
cannot alter our previous soft-pion limit result, therefore,

\beq
\lim_{p_\mu \rightarrow 0} \int_0^\infty ds 
R(Q^2, p^2, s) = 0 \ \ .
\label{Rsftpi}
\eeq

\noindent
It is also convenient to use the following notation for its expansion
in powers of $1/Q^2,$

\beq
R(Q^2, p^2, s) = p^2 h_1(p^2,s) 
  +  {{p^2 h_2(p^2,s)}\over{Q^2 + f(p^2,s)}} 
  + {\cal O} \left({{1}\over{Q^4}} \right) \ \ .
\label{defRemaind}
\eeq

\noindent
We have explicitly introduced a factor of $p^2$ in order to make
sure that this term vanishes in the soft-pion limit as given in
Eq.~(\ref{Rsftpi}). This limit also requires that the functions
$h_i(p^2,s)$ (for $i= 1, 2$) do not have a pole at $p^2 = 0.$ The
above equation is not a formal expansion in orders of $1/Q^2$ since we
have introduced the function $f(p^2,s)$ in the denominator of the
second term. This has been done in order to make
the $Q^2$ integral of this term convergent at low $Q^2.$ The reason
for choosing the above notation will be clear in the following
extraction of the subtraction term contribution to $\delta
m_\pi^{EM}.$

We rewrite the subtraction term contribution,

\beq
\Delta m_\pi^2 (Subtr.) = {{ - 3 \alpha}\over{8 \pi}}
\int_0^\infty \! \! dQ^2 T_1(-Q^2,0) \ \ .
\eeq

\noindent 
The subtraction term including the remainder part, except its
${\cal O} ( 1/Q^4 )$ contributions not present in $h_2$, is

{\everymath{\displaystyle}
\beq
\begin{array}{l} 
T_1(-Q^2,0) \ \ = \ \ -2 + {{2}\over{F_\pi^2}}
  \int_0^\infty \! \! ds 
  \left\{  \rho^R_V(s) {{Q^2}\over{s+Q^2}}
  \right.
\\ \\ \qquad  
  - \rho^R_A(s)
    {{Q^2}\over{s -p^2 +Q^2}} \left( 1 - {{p^2}\over{s}} \right)
  - g_V(s) {{p^2 Q^2}\over{s - p^2 + Q^2}}
\\ \\ \qquad
  \left.
    + p^2 h_1(p^2,s) 
    + {{p^2 h_2(p^2,s)}\over{Q^2 + f(p^2,s)}} 
  \right\} \ \ .   
\end{array}
\label{RSubTrm}
\eeq 
}

\noindent
We expand the above equation in powers of $1/Q^2$ in order to obtain

{\everymath{\displaystyle}
\beq
\begin{array}{l}
T_1(-Q^2,0) \ \ = \ \ 
\\ \\ \ 
  -2 + {{2}\over{F_\pi^2}}
  \int_0^\infty \! \! ds 
  \left[ \rho^R_V(s) 
  - {{s -p^2}\over{s}} \rho^R_A(s) - p^2 g_V(s) + p^2 h_1(p^2,s)
  \right]
\\ \\ \ 
  + \int_0^\infty \! \! ds 
  \left[ s \rho^R_V(s) 
  - {{(s -p^2)^2}\over{s}} \rho^R_A(s) - p^2 (s -p^2) g_V(s) 
  + p^2 h_2(p^2,s) 
  \right] {{1}\over{Q^2}}
\\ \\ \
  + {\cal O} \left( {{1}\over{Q^4}} \right) \ \ .
\end{array}
\label{RSubTrmQ2exp}
\eeq
}

\noindent
The finiteness of $\delta m_\pi^2$ requires the cancellation of the
linear and logarithmic divergences, resulting in the constraints,

{\everymath{\displaystyle}
\begin{eqnarray} 
\int_0^\infty \! \! ds \left\{
  \left[ \rho^R_V(s) - \rho^R_A(s) \right] 
  + p^2 \left[ {{\rho^R_A(s)}\over{s}} - g_v(s) + h_1(p^2,s) 
  \right] \right\} & = & F_\pi^2 \ \ ,
\label{RnuConstr} \\ 
  \int_0^\infty \! \! ds 
  \left\{
   \left[ s \rho^R_V(s) - s \rho^R_A(s) \right] 
   \phantom{ + p^2 \left[ g_V(s) - {{\rho^R_A(s)}\over{s}} \right]
   + p^2 h_2(p^2,s) }  
  \right.
  & & 
\nonumber \\ 
  \left.
  + p^2 \left[ 2 \rho^R_A(s) - s g_V(s) \right]
  + p^4 \left[ g_V(s) - {{\rho^R_A(s)}\over{s}} \right]
  + p^2 h_2(p^2,s) 
  \right\} & = & 0 \ \ .
\label{RQ2Constr}
\end{eqnarray}
}

\noindent
These constraints also reduce to the Weinberg sum rules when we let
$p^2 = 0.$ The role of the functions $h_i(p^2,s)$ (for i = 1, 2) is to
include all other possible contributions. The above constraints must
be satisfied exactly, otherwise the pion EM mass difference would be
divergent.

We can use the Weinberg sum rules, Eqs.~(\ref{W1}) and
(\ref{W2}), to further simplify the previous equations, 

{\everymath{\displaystyle}
\begin{eqnarray} 
p^2 \int_0^\infty \! \! ds
  \left[ {{\rho^R_A(s)}\over{s}} - g_v(s) + h_1(p^2,s) 
  \right] & = & 0 \ ,
\label{RnuConstr1} \\ 
p^2 \int_0^\infty \! \! ds \left\{
  \left[ 2 \rho^R_A(s) - s g_V(s) \right]
  + p^2 \left[ g_V(s) - {{\rho^R_A(s)}\over{s}} \right]
  + h_2(p^2,s) \right\} & = & 0 \ .
\label{RQ2Constr1}
\end{eqnarray}
}

\noindent
We can use solutions available for the functions $\rho^R_A(s)$
\cite{Dono:94a} and $g_V(s)$ to estimate the integrals for the
remainder terms,

{\everymath{\displaystyle}
\begin{eqnarray} 
\int_0^\infty \! \! ds p^2 h_1(p^2,s) & = & 
  \int_0^\infty \! \! ds
  p^2 \left[ g_v(s) - {{\rho^R_A(s)}\over{s}} \right] 
\nonumber \\
& = &  3.0 \times 10^{-4} \ \ ,
\label{h1int} \\ 
\int_0^\infty \! \! ds p^2 h_2(p^2,s) & = &
  \int_0^\infty \! \! ds \left\{
  p^2 \left[ 2 \rho^R_A(s) - s g_V(s) \right]
  + p^4 \left[ g_V(s) - {{\rho^R_A(s)}\over{s}} \right] \right\} 
\nonumber \\
& = &  6.1 \times 10^{-4} \ \ .
\label{h2int}
\end{eqnarray}
}

\noindent
As expected, these values are small when compared to the integrals
involving $\rho^R_V(s)$ which are the larger terms in the constraint 
equations, 

{\everymath{\displaystyle}
\begin{eqnarray} 
\int_0^\infty \! \! ds \rho_V^R(s) & = &  3.83 \times 10^{-2} \ \ ,
\\
\int_0^\infty \! \! ds s \rho_V^R(s) & = &  5.62 \times 10^{-2} \ \ ,
\end{eqnarray}
}

The remainder term $R(Q^2,p^2,s)$ allows us to satisfy the constraints
exactly since it introduces a small correction to the previous
constraint equations. We can proceed to find the subtraction term
contribution by following the steps that we used previously in order
to obtain Eqs.~(\ref{dmpiSublindiv})-(\ref{SubtSpFndmpi})

{\everymath{\displaystyle}
\beq 
\begin{array}{l}
\Delta m_{\pi}^2 (Subtr.) \ \ = \ \ 
\\ \\ \qquad
  {{- 3 \alpha}\over{4 \pi F_\pi^2}}  \int_0^\infty \! \! ds 
  \left\{
   \rho^R_V(s) s \ln s 
   - \rho^R_A(s) {{(s - p^2)^2}\over{s}} 
    \ln \left( s -p^2 \right)
  \right.
\\ \\ \qquad
  \left. 
   - g_V(s) p^2 (s -p^2) 
    \ln \left( s -p^2 \right)
   + p^2 h_2(p^2,s) \ln f(p^2,s) 
   \! \! \!
   \phantom{{{1}\over{1}}} 
   \right\} \ \ .
\label{dmpisubtr}
\end{array}
\eeq
}

\noindent 
Even though the functions $f(p^2,s)$ and $h_2(p^2,s)$ are
undetermined, we have seen in equation~(\ref{h2int}) that their
contributions to the constraint Eq.~(\ref{RQ2Constr}) are small.

\bigskip
\bigskip
\ni
{\it Numerical result}
\bigskip

The total pion electromagnetic mass difference is given by the
addition of the elastic term of Eqs. (\ref{SpFndmpi}), the structure
function terms of Eqs. (\ref{SpFndmpi}) and (\ref{dmpiIntVecW12res}),
and the subtraction constant term of Eq. (\ref{dmpisubtr}).
The results for the narrow-width approximation
and for the corresponding spectral functions is given in
Table~\ref{tab:dmpifin}.

\begin{table}[htbp]
\centering
\caption{$\Delta m_\pi^{EM}$ results.}
\label{tab:dmpifin}
\begin{tabular}{lr@{}c@{}l@{}r@{}c@{}l@{}}
\\
\hline \hline \\
  & \multicolumn{3}{c}{$\mbox{Narrow-width}$} & 
    \multicolumn{3}{c}{$\rho_A^{R}(s), \rho_V^{R}(s) $}\\
  & \multicolumn{3}{c}{\mbox{(MeV)}}  & 
    \multicolumn{3}{c}{\mbox{(MeV)}} \\
\hline \\
Subtr.         & \phantom{0000}4&.&306  & \phantom{0000} 4&.&124 \\
Elastic                     &  0&.&500               &  0&.&500 \\
Str. Fn. $a_1$ int. st.     &  0&.&028               &  0&.&041 \\
Str. Fn. $\omega$ int. st.  & -0&.&127               & -0&.&127 \\
Total calculated            &  4&.&707               &  4&.&538 \\
Experiment                  &  4&.&594               &  4&.&594 \\
\hline  \hline
\end{tabular}
\end{table}

We see from the results that the dominant contribution comes from the
subtraction term, which is largely the effect of vector and
axial-vector resonances, with modest dependence on $p^2 = m_\pi^2$.
The elastic term gives the only other significant contribution. The
modification due to non-zero width is also not large. The overall
result is in excellent agreement with experiment.


\bigskip
\bigskip
\noindent
{\large{\bf 5. The kaon EM mass difference}}
\bigskip
\bigskip

Having set up and tested our methodology for the pion, we now proceed
to the calculation of the kaon electromagnetic mass difference. The
most important effect is that the larger mass of the kaon leads to
kinematic corrections in the various formulae. There are also changes
in the mass, width and couplings of the resonances which we extract
from the data

The kaon calculation is very similar to the pion one described in the
previous section, therefore we will concentrate on the differences
that arise in the kaon case. The kaon counterpart for the Lagrangian
of Eq.~(\ref{Lvector}) expanded in terms of kaon, photon, and
spin-1 resonance fields is

{\everymath{\displaystyle}
\begin{eqnarray}
{\cal L} & = & i e A^\mu ( K^+ \partial_\mu K^- - K^-
\partial_\mu K^+ ) + e^2 A^\mu A_\mu K^+ K^- \ \ ,
\label{LKspzeQED}
\nonumber \\ 
& & - {{e F_V}\over{2}} F^{\mu\nu} \left(
\rho_{\mu\nu}^0 + {{\sqrt{2}}\over{3}} \phi_{\mu\nu} + {{1}\over{3}}
\omega_{\mu\nu} \right)
\nonumber \\
& & + {{e F_V}\over{4 F_K^2}} F^{\mu\nu} \left( \rho_{\mu\nu}^0 +
\sqrt{2} \phi_{\mu\nu} + \omega_{\mu\nu} \right) K^+ K^-
\nonumber \\
& & + {{i G_V}\over{F_K^2}} \left( \rho_{\mu\nu}^0 + \sqrt{2}
\phi_{\mu\nu} + \omega_{\mu\nu} \right) \partial_\mu K^+ \partial_\nu
K^-
\nonumber \\
& & + {{i G_V}\over{F_K^2}} \left( - \rho_{\mu\nu}^0 + \sqrt{2}
\phi_{\mu\nu} + \omega_{\mu\nu} \right) \partial_\mu K^0 \partial_\nu
\overline{K^0} \ \ ,
\label{LKvec}
\nonumber \\
& & - {{i e F_A}\over{2 F_K}} F^{\mu\nu} \left(
K_{1_{\mu\nu}}^- K^+ - K_{1_{\mu\nu}}^+ K^- \right) \ \ ,
\label{LKaxialvec}
\end{eqnarray}
}

\ni
where we have used ideal mixing for the vector meson resonances.

The major difference between the kaon and pion Lagrangians -
Eqs.~(\ref{Lvector}) and (\ref{LKaxialvec}) respectively - is that in the
kaon case all the three nonet vector resonances contribute. Another difference is
that there is an elastic contribution to the neutral kaon self-energy. This
contribution vanishes in the ideal mixing approximation together with the limit
where all the vector resonance masses are equal.

The contribution to the kaon Compton scattering amplitude given by the Feynman
diagrams of Figs. \ref{fig:CSFeynDiag}.a and \ref{fig:CSFeynDiag}.b is

{\everymath{\displaystyle}
\beq
\begin{array}{l}
T_{\mu\nu}^{(1)(K)}(q^2,\pq) \ \ = \ \ 
 - 2 D_{1_{\mu\nu}} 
 + 4 m_K^2 D_{2_{\mu\nu}}
\\ \\  \qquad
  \times \ \left[ ( G_{K^+}(q^2) )^2 - ( G_{K^0}(q^2) )^2 \right]
  \left( {{1}\over{m_K^2 - (p+q)^2 -i \epsilon}} 
    + ( q \rightarrow -q ) \right) \ \ ,
\end{array}
\label{KchCSA1}
\eeq
}

\ni
where

\beq
G_{K^{+,0}}(q^2) \ \ \equiv \ \ \int \! \! du  
  {{u}\over{u - q^2}} \delta_{K^{+,0}}(u) \ \ ,
\eeq

\noindent 
and

\beq
\delta_{K^{+,0}}(u) \ \ \equiv \ \ 
 \pm {{1}\over{2}} \delta (u -m_{\rho}^2)
   + {{1}\over{3}} \delta (u - m_{\phi}^2 ) 
   + {{1}\over{6}} \delta (u - m_{\omega}^2 ) \ \ .
\label{defdelK+}
\eeq

\ni
We have subtracted the neutral kaon contribution in order to be able to use this
equation in the following kaon EM mass difference formulas.

The vector seagull contribution, Fig.~\ref{fig:CSFeynDiag}.c,  is

\beq
T_{\mu\nu}^{(2)(K)}(q^2,\pq) = 
  - 2 {{F_V^2}\over{F_K^2}} \int \! \! du
    {{q^2}\over{u - q^2}} \delta_{K^+}(u) D_{1_{\mu\nu}} \ \ .
\label{KCSA2}
\eeq

\noindent
Finally, the axial-vector resonance intermediate state contribution,
Fig.~\ref{fig:CSFeynDiag}.d, is

{\everymath{\displaystyle}
\beq
\begin{array}{l}
T^{(3)(K)}_{\mu \nu}(q^2,\pq) =  
\\ \\ 
  \qquad
  {{{F_A^K}^2}\over{F_K^2 m_{K_{1_A}}^2}}
  \left( { {\left( \pq + q^2 \right)^2 
  +q^2 \left( m_{K_{1_A}}^2 -(p+q)^2\right)}
    \over{m_{K_{1_A}}^2 -(p+q)^2 -i \epsilon}}
  + ( q \rightarrow -q )   \phantom{ {{1}\over{1}} }   
   \right) {D_1}_{\mu \nu} 
  \\ \\
  \qquad
  + {{{F_A^K}^2}\over{F_K^2 m_{K_{1_A}}^2}}
 \left( {{- m_K^2 q^2} \over{m_{K_{1_A}}^2 - (p+q)^2 -i \epsilon }} 
      + (q \rightarrow -q)  \right) {D_2}_{\mu \nu} \ \ .
\end{array}
\label{KCSA3}
\eeq
}

\noindent
The axial vector intermediate state for the kaon is less
straightforward than for the pion since the axial-vector meson
$K_{1_A}$ is an ill-determined mixture of the physical states $K_1(1270)$
and $K_1(1400)$ \cite{PDG:94}. We will treat this issue later when we estimate
the spectral function for the axial-vector intermediate state.

The breakdown into structure functions and subtraction term of the Compton
scattering amplitude given by the above three terms 
$T_{\mu\nu}^{(i)(K)},$ (for $i = 1$ to $3$), 
in the spectral function representation is

{\everymath{\displaystyle}
\begin{eqnarray}
T_1(-Q^2,0) & = & -2 + {{2}\over{F_K^2}} 
  \int_0^\infty \! \! ds \rho_V^K(s) {{Q^2}\over{s+Q^2}}
\nonumber \\
& &  - {{2}\over{F_K^2}} \, 
  \int_0^\infty \! \! ds \, \rho_V^K(s) {{Q^2}\over{s -p^2 +Q^2}}
  \left( 1 - {{p^2}\over{s}} \right) \ \ ,
\nonumber \\   
W_1(-Q^2,\nu) & = & {{1}\over{F_K^2}} 
  \int \! \! ds \, {{\rho_A^K(s)}\over{s}} \, 
  ( \nu -Q^2 )^2 \, \tilde{K}(\nu,s) 
  \delta (s -p^2 +Q^2 - 2 \nu ) \ \ ,
\nonumber \\   
W_2(-Q^2,\nu) & = & 4 m_K^2 \delta (Q^2 - 2 \nu ) 
  \left[ (G_{K^+}(-Q^2))^2 - (G_{K^0}(-Q^2))^2 \right]
\nonumber \\
& & + {{1}\over{F_K^2 \, s}} \int \! \! ds \,
  {{\rho_A^K(s)}\over{s}} \, p^2 Q^2  \tilde{K}(\nu,s) 
  \delta (s -p^2 +Q^2 -2 \nu ) \ \ , \label{KSpFnTW}
\end{eqnarray}
}

\noindent 
where $\nu = \pq = m_K q_0$, and $p^2 = m_K^2$. We have also introduced the
convergence factor

{\everymath{\displaystyle}
\begin{eqnarray}
\tilde{K}(\nu,s) & = & 
  \int \! \! du 
  \left( {{u}\over{u +2 \nu}} \right)^4
  \left( 1 + \tilde{\eta}(u) {{2 \nu}\over{s}} \right) 
  \delta_{K^+}(u) \ \ ,
\label{defKtildefactr}
\end{eqnarray}
}

\noindent
where

\beq
\tilde{\eta}(u) =  {{s}\over{s - p^2}} 
  \left[ \left( 1 + {{s - p^2}\over{u}} \right)^4 -1 \right] \ \ ,
\eeq

\noindent
and $p^2 = m_K^2 \ .$

The definition of the kaon vector spectral function is

\beq
\rho_V^K(s) \equiv {{1}\over{2}} \rho_\rho^R(s) 
+ {{1}\over{2}} \rho_\phi^R(s) + {{1}\over{6}} \rho_\omega^R(s) \ \ ,
\eeq

\ni
where $\rho_\rho^R(s)$ is the spectral function introduced in
Fig.~\ref{fig:RhoARVRC}, and the other two are due to the $\phi$ and
$\omega$ intermediate states. For these last two it is appropriate to
use the narrow-width approximation.

We also include the VP$\gamma$ vertices in the same way as we did for the
pions. The effective Lagrangian that we use for the
$K^*K\gamma$ vertex is

\beq
{\cal L} = e \, {{\sqrt{R_V}}\over{2}} \, 
  \tilde{K}(\nu,m_{K^*}^2) \, \epsilon^{\mu \nu \alpha \beta} \,
  F_{\mu \nu} V_{\alpha} \partial_\beta K \ \ ,
\label{LKIntVec}
\eeq 

\ni
The subtraction term and structure functions that follow from the
above Lagrangian, including the $K^0$ functions with an extra minus
sign to be able to insert them directly in the kaon EM mass difference
formula are

{\everymath{\displaystyle}    
\begin{eqnarray}              
T_1 (-Q^2,0) & = & {{2}\over{F_K^2}} 
   \int_0^\infty \! \! ds g^K_V(s)
  {{p^2 Q^2}\over{s - p^2 + Q^2}} \ \ ,
\nonumber \\
W_1 (-Q^2,\nu) & = & {{1}\over{F_K^2}} 
  \int_0^\infty \! \! ds g^K_V(s)
  (\nu^2 + p^2 Q^2) \ \tilde{K}(\nu,s) 
  \delta(s - p^2 + Q^2 - 2 \nu) \ \ ,
\nonumber \\
W_2 (-Q^2,\nu) & = & {{1}\over{F_K^2}} 
  \int_0^\infty \! \! ds g^K_V(s)
  p^2 Q^2 \ \tilde{K}(\nu,s)
  \delta(s - p^2 + Q^2 - 2 \nu) \ \  .
\label{KIntVecTW}
\end{eqnarray}
}                             

\ni 
The narrow-width approximation is justified in this case due to the small width
of the $K^*$ intermediate states. Therefore, we use the following definition,

\beq
g^K_V(s) \ \ \equiv \ \ 
   H_{{K^*}^0}^2 \: \delta(s - m_{{K^*}^0}^2) -
   H_{{K^*}^+}^2 \: \delta(s - m_{{K^*}^+}^2) \ \ .
\eeq

Unlike in the pion case, there is an intermediate vector meson contribution for
the neutral kaon as well as for the charged kaon, in the SU(3) limit. For the
pions, the intermediate $\rho$ meson contribution canceled in the SU(2)
limit, (since charged and neutral $\rho$ couplings become equal), leaving only the
intermediate $\omega$ and $\phi$ contributions to the neutral pion Compton
scattering amplitude.

We determine the $K^*$ couplings from the radiative decays $K^*
\rightarrow K \gamma \ ,$

{\everymath{\displaystyle}    
\begin{eqnarray}  
R_{K^{*^+}} & = & {{24}\over{\alpha}}
    {{m_{K^{*^+}}^3 }\over{(m_{K^{*^+}}^2 - m_K^2)^3}} 
    \ \Gamma_{K^{*^+} \rightarrow K^+ \gamma}
  \ = \ 0.70 \pm  0.06 \ \mbox{GeV}^{-2} \ \ ,
\nonumber \\
R_{K^{*^0}} & = & {{24}\over{\alpha}}
    {{m_{K^{*^0}}^3}\over{(m_{K^{*^0}}^2 - m_K^2)^3}} 
    \ \Gamma_{K^{*^0} \rightarrow K^0 \gamma}
  \ = \ 1.61 \pm 0.14 \ \mbox{GeV}^{-2} \ \ ,
\nonumber 
\end{eqnarray}
}

\ni
from which we obtain the values for

{\everymath{\displaystyle}    
\begin{eqnarray}  
H_{K^{*^+}} & = & F_K \sqrt{R_{K^{*^+}}} \ \ =  \ \ 
  0.093 \pm 0.004 \ \ ,
\nonumber \\
H_{K^{*^0}} & = & F_K \sqrt{R_{K^{*^0}}} \ \ = \ \ 
  0.143 \pm 0.006 \ \ .
\end{eqnarray}
}

In the kaon case, the mass difference need not be finite because there
can be divergences which are absorbed into the renormalized masses of
the up and down quarks. However this effect is relatively small
because it is proportional to $\alpha m_u$ or $\alpha m_d$ compared to
the dominant electromagnetic mass-shift which is simply of order
$\alpha$. We assume that the renormalization of the up and down quark
masses has been carried out, although the precise renormalization
prescription is hard to define because of the small size of this
effect. The remaining electromagnetic effects are finite.

We can now determine the full high $Q^2$ constraints for a finite kaon EM
self energy given by combining Eqs.~(\ref{KSpFnTW}) and (\ref{KIntVecTW}),
and including the remainder terms introduced in
Eq.~(\ref{defRemaind}),

{\everymath{\displaystyle}
\begin{eqnarray} 
\int_0^\infty \! \! ds \left\{
  \left[ \rho^K_V(s) - \rho^K_A(s) \right] 
  + p^2 \left[ {{\rho^K_A(s)}\over{s}} - g^K_v(s) + h_1^K(p^2,s) 
  \right] \right\} & = & F_K^2 \ \ ,
\label{RQ0ConstrK} \\ 
  \int_0^\infty \! \! ds 
  \left\{
   \left[ s \rho^K_V(s) - s \rho^K_A(s) \right] 
   \phantom{ + p^2 \left[ g^K_V(s) - {{\rho^K_A(s)}\over{s}} \right]
   + p^2 h_2^K(p^2,s) }  
  \right.
  & & 
\nonumber \\ 
  \left.
  + p^2 \left[ 2 \rho^K_A(s) - s g^K_V(s) \right]
  + p^4 \left[ g^K_V(s) - {{\rho^K_A(s)}\over{s}} \right]
  + p^2 h_2^K(p^2,s) 
  \right\} & = & 0 \ \ ,
\label{RQ2ConstrK}
\end{eqnarray}
}

\noindent
where $p^2 = m_K^2$. These equations are similar to the ones
encountered in the pion case. 


Using the above constraints and all available data for the vector
spectral function, $\rho_V^R(s)$ and the narrow-width approximation
for $g_V(s)$, and the $\omega$, $\phi$, $\omega(1420)$,
$\omega(1600)$, and $\phi(1680)$ resonances we obtain for $h_1 = h_2 = 0$

{\everymath{\displaystyle}
\begin{eqnarray} 
\int \! ds \rho_{A}^K(s) \ \left(1 - {{p^2}\over{s}} \right)   
  & = &  0.0170(12) \ \mbox{GeV}^2 \ \ , 
\label{NumConstr1} \\
\int \! ds \, s \rho_{A}^K(s) \  \left(1 - {{p^2}\over{s}} \right)^2 
  & = &  0.0406(16) \ \mbox{GeV}^4 \ \ ,   
\label{NumConstr2} 
\end{eqnarray}
}

\ni 
where $p^2 = m_K^2$.

We try first the narrow-width approximation to the axial-vector spectral function

{\everymath{\displaystyle}
\begin{eqnarray}
\rho_A^{K^{(NW)}}(s) & = & 
  {{F^K_{A}}^{2}}_{(NW)} \left[ \cos^2 \theta_{K} \delta ( s - m^2_{K_1(1400)} ) 
  \right.
\nonumber \\
  & & \left. + \sin^2 \theta_{K} \delta ( s - m^2_{K_1(1270)} )
  + R_K^2 \delta ( s - m^2_{K_1(HR)} ) \right] \  \ .
\label{defNWSpFnK}
\end{eqnarray}
}

\ni 
This parametrization includes the $K_1(1270)$ and $K_1(1400)$ resonances as well
as a higher mass resonance $K_1(HR)$. The input parameters to $\rho_A^K$ for this
calculation are

{\everymath{\displaystyle}
\begin{eqnarray}
\theta_K  =  {{\pi}\over{4}} 
\ \ \ \mbox{and} \ \ \ 
m_{K_1(HR)}   =  2.0 \ \mbox{GeV} \ \ .
\label{K1param}
\end{eqnarray}
}

\ni
These choices are sensible but arbitrary. They fix the values of
$F_A^K$ and $R_K$ through the constraint Eqs.~(\ref{NumConstr1}) and
(\ref{NumConstr2}). The obtained values for $F_A^K$ and $R_K$ show a
sizeable dependence on the choice of $m_{K_1(HR)}$.  The results for
the different contributions to $\delta m_K^{EM}$ obtained by this
narrow-width approximation are given in Table~\ref{tab:dmkfin}. In
particular we find that the subtraction term contribution is very
large. However, this contribution varies from $\delta m_K^{(Subtr.)}
\sim 2.3$ MeV for $m_{K_1(HR)} = 1.8$ GeV to $\delta m_K^{(Subtr.)}
\sim 3.1$ MeV for $m_{K_1(HR)} = 2.4$ GeV.  These numerical results
only constitute a very rough estimate. This is already indicated by
the large dependence in $m_{K_1(HR)}$, and once again, it involves the
narrow-width approximation for broad resonances.

There is another constraint on the axial-vector spectral function due
to $\tau$ decay

\beq 
BR(\tau \rightarrow \nu_\tau K_1 \rightarrow \nu_\tau K \pi \pi) =
E_\tau \int_{0}^{m_\tau^2} \! ds \rho_{A}^K(s)  
  \left(1 - {{s}\over{m_\tau^2} } \right)^2
  \left( 1 + {{2 s}\over{m_\tau^2}} \right)  \ \ ,
\label{cstr3} 
\eeq

\ni 
where 

{\everymath{\displaystyle}
\beq
E_{\tau} =  
{{G_{\mu}^2 m_{\tau}^3 |V_{us}|^2}\over{8 \pi \Gamma_\tau}} 
\ \ =  \ \  0.6633 \ \mbox{GeV}^{-2} \ \ .
\eeq
}

The data for the tau lepton decay, $\tau \rightarrow \nu_{\tau} K \pi
\pi \:,$  gives the branching ratios \cite{Helt:94},
 
\begin{eqnarray} 
BR(\tau^- \rightarrow \nu_\tau K^- \pi^+ \pi^-) 
 & = & ( 0.40 \pm 0.09 ) \%  \ \ , \\
BR(\tau^- \rightarrow \nu_\tau \bar{K}^0 \pi^- \pi^0) 
 & = & ( 0.41 \pm 0.07 ) \%  \ \  , \\
BR(\tau^- \rightarrow \nu_\tau K^- \pi^0 \pi^0) 
 & = & ( 0.09 \pm 0.03 ) \%  \ \  , \\
BR(\tau \rightarrow \nu_\tau K \pi \pi)
 & = & ( 0.90 \pm 0.12 ) \%  \ \ .
\label{exptaubr}
\end{eqnarray}

\ni
The last branching ratio is the sum of the three different decay channels with
the uncertainties added in quadrature. Even though we expect these branching
ratios to be dominated by the axial-vector channels specially the $K_1(1270)$
and the $K_1(1400)$, there should also be a contribution due to the vector
resonance $K^*(1410)$ \cite{Asto:87}. This resonance will contribute through
the decay process $K^*(1410) \rightarrow K^*(892) \pi \rightarrow K \pi \pi$.
The branching ratio for $K^*(1410) \rightarrow K^*(892) \pi$ is greater than
40\% at 95\% confidence level \cite{PDG:94}, and $BR(K^*(892) \rightarrow
K \pi) \sim 100\%$ . Therefore, the tau branching ratio into the strange
axial-vector channels should be somewhat lower than stated in
Eq.~(\ref{exptaubr}).

We obtain the shape of $\rho_A^K(s)$, up to $m_{K\pi\pi} = 2.1$ GeV,
from diffractive production experimental data obtained by the ACCMOR
collaboration in 1981 \cite{Daum:81}. The data was extracted from
200,000 examples of the reaction $K^- p \rightarrow K^- \pi^- \pi^+
p$. The intensity for the $1^+$ channel gives us the shape of
$\rho_A^K(s)$.

We add a high energy tail to the data, up to $m_{K\pi\pi} = 2.54$ GeV,
which decreases quadratically. The constraint equations favor this
quadratic choice instead of other simple parametrizations.
The final normalization of the spectral function is
obtained by enforcing the constraint Eqs.~(\ref{NumConstr1}) and
(\ref{NumConstr2}). If we define

\beq
\rho^{K}_A(s) \ \ \equiv \ \
{F^K_A}^{2} \ \ \bar{\rho}_{A}(s) \  \ , 
\quad
\mbox{where}  
\quad
\int \! ds \bar{\rho}_{A}(s) = 1
\ \ , 
\eeq

\begin{figure}[htbp]
\centerline{\psfig{figure=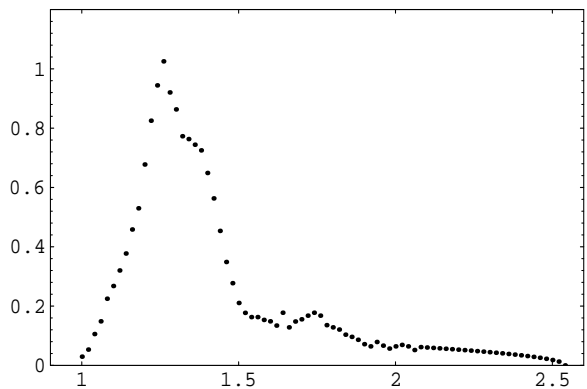,height=2.00in,width=3.00in}}
\caption[] 
{ \sf $\bar{\rho}^{(i)}(s)$ vs.
       $m_{K\pi\pi}$ (GeV), for  $i = 1$ to $3$ .
}
\label{fig:rhobar}
\end{figure}

\ni
we obtain $F^K_A = 0.144$ GeV, and the $\bar{\rho}_{A}(s)$ given in
Fig.~\ref{fig:rhobar}. 

The choice of $\rho_A^{K}$ results in an estimate for
the tau branching ratio for the $K^*(1410)$ vectorial resonance, 

\beq
BR(\tau \rightarrow \nu_\tau K^*(1410))  \ \ = \ \ 
( 0.46 \pm 0.13 ) \%  \ \ . 
\eeq

\ni
this value could be extracted from data by doing an angular momentum
analysis of the $K \pi \pi$ final state.

The final results for the different contributions
to $\Delta m_K$ are given in Table~\ref{tab:dmkfin}. We estimate the
uncertainty of the total kaon EM mass difference to be $\sigma (\Delta
m_K) \sim 0.6$ MeV.

\begin{table}[htbp]
\centering
\caption{$\Delta m_K^{EM}$ results.}
\label{tab:dmkfin}
\begin{tabular}{lr@{}c@{}l@{}r@{}c@{}l@{}}
\\
\hline \hline \\
  & \multicolumn{3}{c}{$\mbox{Narrow-width}$} & 
    \multicolumn{3}{c}{$\rho_A^{K}(s), \rho_V^{K}(s) $}\\
  & \multicolumn{3}{c}{\mbox{(MeV)}}  & 
    \multicolumn{3}{c}{\mbox{(MeV)}} \\
\hline \\
Subtr.        & \phantom{0000} 2&.&56  & \phantom{0000} 1&.&80 \\
Elastic                     &  0&.&92                &  0&.&92 \\
Str. Fn. $K_1$ int. st.     &  0&.&05                &  0&.&07 \\
Str. Fn. $K^*$ int. st.     & -0&.&18                & -0&.&18 \\
Total calculated            &  3&.&35                &  2&.&61 \\
Dashen                      &  1&.&27                &  1&.&27 \\
\hline  \hline
\end{tabular}
\end{table}

Our result is about 100\% greater than Dashen's result. This result is
in better agreement with earlier references
\cite{Bijn:93,Dono:93a,Gass:85b}, and with the recent investigation
\cite{Bijn:96}, but in disagreement with Baur and Urech
\cite{Baur:96}. Given the uncertainty of our result, we feel
more comfortable by saying that we find a modification of Dashen's
theorem of between 160\% and 240\% .



\bigskip
\bigskip
\noindent
{\large{\bf 6. Conclusions}}
\bigskip
\bigskip

The calculation of nonleptonic amplitudes is in general one of
the most difficult tasks for analytic strong interaction techniques.
The elecromagnetic mass differences of the pseudoscalar mesons seems
to us to be the most favorable case to attempt a controlled
calculation. There turn out to be several favorable circumstances that
help in this endeavor. As we have exploited above, the  relevant
current-current products have several connections to known
phenomenology, and have important constraints due to the long distance
chiral behavior and the short distance properties of $QCD$.

The calculation of the known pion mass difference was quite
successful. It turns out that intermediate mass scales (around 1
GeV) are the most important for this matrix element, and these are
well represented by resonance contributions. In fact this structure is
already visible in the old calculation in the soft pion limit given by
Das et al. where the vector and axial vector spectral functions
determine the mass difference in the chiral limit. There are
calculable corrections and even new diagrams that come in as one
includes a non-zero pion mass, but the pion mass is still small enough
that one does not change the general anatomy of the matrix element.

In the case of the kaon mass difference, the experimental result is
not known. We find a large deviation from the prediction of Dashen's
theorem, which is valid in the limit of massless kaons. While the
magnitude of this effect is larger than most $SU(3)$ breaking effects
in chiral calculations, we stress that its origin is in reasonably
well-known and mundane effects, and does not represent any breakdown
of chiral symmetry. The main effect seems to be the kaon mass in the
propagator of the Born diagram, which hence is a rather long distance
effect, while the remaining dependence comes from the known shift in
resonance masses due the the strange quark mass. This mass difference
is important for the extraction of the $u-d$ quark mass difference.





\bigskip
\bigskip
\noindent
{\large{\bf Appendix}}
\bigskip
\bigskip

The notation used for the {\cal O}($E^2$) chiral terms in the
Lagrangian of Eq.~(\ref{Lchires}) is

{\everymath{\displaystyle}
\begin{eqnarray}
U    & = & exp \left( i \sqrt{2} \Phi / F \right) \ \ , 
\nonumber \\
\Phi & = & 
\left(  
\begin{array}{ccc}  
{{\pi^0}\over{\sqrt{2}}} + {{\eta_8}\over{\sqrt{6}}} & \pi^+ & K^+ \\
\pi^- & - {{\pi^0}\over{\sqrt{2}}} + {{\eta_8}\over{\sqrt{6}}} & K^0 \\
K^- & \bar{K^0} & - {{2}\over{\sqrt{6}}} \eta_8
\end{array} 
\right) \ \ , 
\nonumber \\
\chi & = & 2 B_0 
\left(  
\begin{array}{ccc}
m_u &  0   &  0   \\
0   &  m_d &  0   \\
0   &  0   &  m_s
\end{array}
\right) \ \ , 
\nonumber \\
D_\mu U & = & \partial_\mu U - i (v_\mu + a_\mu) U + i U (v_\mu - a_\mu) 
\ \ , \label{defDU}
\end{eqnarray}
}
\noindent 
where $v^\mu$, $a^\mu$ are the external fields. In order to include EM one
needs to define 

\beq
a_\mu = 0 \, , \quad v_\mu = e Q A_\mu \ \ ,
\eeq

\ni
where $A_\mu$ is the photon field, and should not be confused with the
axial-vector antisymmetric tensor field which has two Lorentz indices,
$A_{\mu\nu}.$ $Q$ is the quark charge matrix, for the $u,$ $d$ and $s$
quarks,

\beq
Q  =  
\left(  
\begin{array}{ccc}  
{{2}\over{3}} & 0 & 0 \\
0 & - {{1}\over{3}} & 0 \\
0 & 0 & - {{1}\over{3}} 
\end{array} 
\right) \ \ . \\
\eeq

The notation used for Lagrangian containing the chiral couplings of the vector
and axial-vector meson resonances, Eq.~(\ref{Lchires}) is

{\everymath{\displaystyle}
\begin{eqnarray}
u_\mu & = & i u^\dagger D_\mu U u^\dagger \ \ = \ \ u_\mu^\dagger \ \ ,
\nonumber \\
f_\pm^{\mu\nu} & = & u F_L^{\mu\nu} u^\dagger \pm 
  u^\dagger F_R^{\mu\nu} u \ \ , 
\nonumber \\
F^{\mu\nu}_{R,L} & = & \partial^\mu (v^\nu \pm a^\nu) -
                   \partial^\nu (v^\mu \pm a^\mu) -
                   i [ v^\mu \pm a^\mu, v^\nu \pm a^\nu] \ \ , 
\nonumber \\
V_{\mu\nu} & = & 
\left(  
\begin{array}{ccc}  
{{\rho^0}\over{\sqrt{2}}} + {{\omega_8}\over{\sqrt{6}}} & \rho^+ & K^{*+} \\
\rho^- & - {{\rho^0}\over{\sqrt{2}}} + {{\omega_8}\over{\sqrt{6}}} & K^{*0} \\
K^{*-} & \bar{K^{*0}} & - {{2}\over{\sqrt{6}}} \omega_8
\end{array} 
\right)_{\mu\nu} \ \ , \nonumber \\
\nonumber \\
A_{\mu\nu} & = & 
\left(  
\begin{array}{ccc}  
{{a_1^0}\over{\sqrt{2}}} + {{f_1}\over{\sqrt{6}}} & a_1^+ & K_1^+ \\
a_1^- & - {{a_1^0}\over{\sqrt{2}}} + {{f_1}\over{\sqrt{6}}} & K_1^0 \\
K_1^- & \bar{K_1^0} & - {{2}\over{\sqrt{6}}} f_1
\end{array} 
\right)_{\mu\nu} \ \ ,
\end{eqnarray}
}

{\everymath{\displaystyle}
\begin{eqnarray}
\nabla_{\lambda} R_{\mu\nu} & = & \partial_{\lambda} R_{\mu\nu} + 
  [ \Gamma_{\lambda} , R_{\mu\nu} ] \ \ , 
\nonumber \\
R_{\mu\nu} & = & V_{\mu\nu}, A_{\mu\nu} \ \ ,
\nonumber \\
\Gamma_{\mu} & = & {{1}\over{2}} \left\{ 
  u^\dagger [ \partial_\lambda - i (v_\lambda + a_\lambda) ] u 
  +  u [ \partial_\lambda - i (v_\lambda - a_\lambda) ] 
     u^\dagger \right\} \ \ .
\end{eqnarray}
}

From the kinetic terms of the Lagrangian in equation~(\ref{Lchires})
one derives the free propagator for the antisymmetric tensor field
representation~\cite{Ecke:89a},

{\everymath{\displaystyle}
\beq
\begin{array}{l}
\left< 0|T R_{\mu\nu}(x) R_{\rho\sigma}(y) |0 \right> \ \ = \ \ 
  {-i\over{M^2}} \int {{d^4 k}\over{2 \pi^4}} \:
  {{e^{-i(x-y)}}\over{M^2 - k^2 - i \epsilon}}
\\ \\ \qquad  
  \times \left[
  g_{\mu\rho} g_{\nu\sigma}
  \left( M^2 - k^2 \right) + g_{\mu\rho} k_\nu k_\sigma 
  - g_{\mu\sigma} k_\nu k_\rho - \left( \mu \leftrightarrow \nu \right)
  \right] \ \ ,
\end{array}
\label{VAprop}
\eeq
}

\noindent 
where the normalization is given by

\beq
\left< 0| R_{\mu\nu} | R(\epsilon,p) \right> = {- i \over M} 
\left[ p_\mu \epsilon_\nu(p) - p_\nu \epsilon_\mu(p) \right] \ \ .
\eeq






\vfill \eject

\end{document}